\documentclass[12pt,journal,draftcls,onecolumn]{IEEEtran}
\usepackage{amsmath,epsfig,times}
\usepackage{latexsym,epsf,graphicx,psfrag,amssymb,cite}
\usepackage{cite}
\usepackage{amsmath,tikz}
\usepackage{amssymb}
\usepackage{epsfig}
\usepackage{graphicx}
\usepackage{subfigure}
\usepackage{multirow}
\usepackage{array}
\usepackage{enumerate}
\begin{document}

\title{On the Performance of  Selection Cooperation with Imperfect Channel Estimation}
\author{
Mehdi~Seyfi,~\IEEEmembership{Student Member,~IEEE,}
Sami~(Hakam)~Muhaidat*,~\IEEEmembership{Member,~IEEE,}
and Jie~Liang,~\IEEEmembership{Member,~IEEE}
\thanks{The work of S. Muhaidat was supported in part by the Natural Sciences and Engineering Research Council (NSERC) of Canada under grant RGPIN372049. The work of Jie Liang was supported in part by NSERC of Canada under grants RGPIN312262-05, EQPEQ330976-2006, and
STPGP350416-07. The material in this paper was
presented in part at the 2010 IEEE Vehicular Technology Conference, {\it VTC'10}, Taipei, Taiwan.}
\thanks{M. Seyfi, S. Muhaidat and J. Liang are with the School of Engineering
Science, Simon Fraser University, Burnaby, BC, V5A 1S6, Canada. Phone:
778-782-7376. Fax: 778-782-4951. E-mail: msa119@sfu.ca, muhaidat@ieee.org, jiel@sfu.ca.}
\thanks{}
\thanks{$^*$Corresponding author
}}
\maketitle

\markboth{IEEE TRANSACTIONS ON Wireless Communication, SUBMITTED DEC. 2010.}{M. Seyfi \lowercase{\textit{et al.}}: {On the Performance of Selection Cooperation with Imperfect Channel Estimation}}

\def\arg{{\rm arg}}
\def\Re{{\rm Re}}
\def\Im{{\rm Im}}
\def\bA{{\bf A}}
\def\ba{{\bf a}}
\def\bB{{\bf B}}
\def\bb{{\bf b}}
\def\bC{{\bf C}}
\def\bc{{\bf c}}
\def\bD{{\bf D}}
\def\bd{{\bf d}}
\def\bE{{\bf E}}
\def\be{{\bf e}}
\def\bF{{\bf F}}
\def\bG{{\bf G}}
\def\bg{{\bf g}}
\def\bH{{\bf H}}
\def\bh{{\bf h}}
\def\bI{{\bf I}}
\def\bJ{{\bf J}}
\def\bK{{\bf K}}
\def\bk{{\bf k}}
\def\bM{{\bf M}}
\def\bL{{\bf L}}
\def\bN{{\bf N}}
\def\bn{{\bf n}}
\def\bP{{\bf P}}
\def\bp{{\bf p}}
\def\bQ{{\bf Q}}
\def\bq{{\bf q}}
\def\bR{{\bf R}}
\def\br{{\bf r}}
\def\bS{{\bf S}}
\def\bs{{\bf s}}
\def\bT{{\bf T}}
\def\bt{{\bf t}}
\def\bU{{\bf U}}
\def\bu{{\bf u}}
\def\bV{{\bf V}}
\def\bv{{\bf v}}
\def\bW{{\bf W}}
\def\bw{{\bf w}}
\def\bX{{\bf X}}
\def\bx{{\bf x}}
\def\bY{{\bf Y}}
\def\by{{\bf y}}
\def\bZ{{\bf Z}}
\def\bz{{\bf z}}

\def\rE{{\rm E}}
\def\cM{{\mathcal{M}}}
\def\cC{{\mathcal{C}}}
\def\cD{{\mathcal{D}}}
\def\cE{{\mathcal{E}}}
\def\cI{{\mathcal{I}}}
\def\cN{{\mathcal{N}}}
\def\cS{{\mathcal{S}}}
\def\NOP{{\sqrt{{\it {\mathcal{E}_{_{si}}}}|\hat{h}_{_{si}}|^2+{\it N_0}}}}
\def\NOB{{\sqrt{{\mathcal{E}_{_{si}}}|\hat{h}_{_{si}}|^2+{\mathcal{E}_{_{id}}}|\hat{h}_{_{id}}|^2+N_0}}}
\def\no{{\nonumber}}

\begin{abstract}
In this paper, we investigate the performance of selection cooperation in the presence of imperfect channel estimation. In particular, we consider a cooperative scenario with multiple relays and amplify-and-forward protocol over frequency flat fading channels. In the selection scheme, only the ``best'' relay which maximizes the  effective signal-to-noise ratio (SNR) at the receiver end is selected. We present lower and upper bounds on the effective SNR and derive closed-form expressions for the average symbol error rate (ASER), outage probability and average capacity per bandwidth of the received signal in the presence of channel estimation errors. A simulation study is presented to corroborate the analytical results and to demonstrate the performance of relay selection with imperfect channel estimation.

\end{abstract}
\newpage
\section{Introduction}
\PARstart{T}he promise of high spectral efficiency and capability of providing great capacity improvements in a wireless fading environment,
as reported by \cite{Telatar1999} and \cite{Foschini1995}, has led to widespread interest in multi-input multi-output (MIMO) communications.
However, due to size, cost, and/or hardware limitations, a wireless device may not be able to support multiple transmit antennas.
Cooperative diversity was proposed as  an alternative to MIMO systems. It has been demonstrated  that cooperative diversity  provides an effective way of improving spectral and power efficiency of the wireless networks  \cite{Laneman2004,Nabar2004}. The main idea behind cooperative diversity is that in a wireless environment, the signal transmitted by the source ($S$) is overheard by other nodes, which can be defined as ``relays'' \cite{Giannakis2005}. The source and its partners can then jointly process and transmit their information, thereby creating a ``virtual antenna array'', although each of them is equipped with only one antenna. It is shown in \cite{Anghel2004,Adve2007,Adve2006,Beres2008,Ikkic2009,Ikki2008,Ikki2009,Ikki2010,Jaffarkhani2009} that cooperative diversity networks can achieve a diversity order equal to the number of paths between the source and the destination, however, the need of transmitting the symbols in a time division multiplexing (TDMA) fashion limits the improvement in capacity. Additionally due to the power allocation constraints, multiple relay deployment is not economic.
Relay selection then was proposed to alleviate the loss in spectral efficiency caused by multiple relay schemes and also to moderate the power allocation constraints.

Recently, there have been considerable research efforts on the performance analysis of cooperative diversity including the derivation of closed-form formulas for the average symbol error rate (ASER) \cite{Giannakis2005,Anghel2004,Adve2006,Ikkic2009,Ikki2008,Ikki2009,Ding},  the outage probability \cite{Adve2007,Beres2008,Ikkic2009,Ikki2010,Bletsas2006}, and the average capacity (ergodic capacity) \cite{Adve2007,Ikkic2009,Ikki2008,Poor2009}. Most of the current works, however,  assume the availability of perfect channel state information (CSI) at the destination ($D$) terminal. Channel estimation for relay communication has also been studied in literature  as in\cite{Gedik2008,Gedik2009}.

\emph{Related work and contributions:} The impact of channel estimation error on the ASER performance of distributed space time block coded (DSTBC) systems has been investigated in \cite{Cheng2005}, assuming the amplify-and-forward protocol. Building upon a similar set-up, Gedik and Uysal \cite{Gedik} have extended the work of \cite{Cheng2005} to a system with $M$ relays, assuming the amplify-and-forward protocol. In \cite{Han2009}, the symbol error analysis is investigated for the same scenario as in \cite{Gedik}. In this paper, we investigate the effect of channel estimation error on the ASER, outage probability and capacity of a cooperative diversity system with relay selection.
 We first introduce our system model in Sec. \ref{sec system model}. In Sec. \ref{Sec_Error model}, we explain the Gaussian error model introduced by \cite{Leung2003} and use it in the selection scenario to obtain the maximum ratio combiner (MRC) output SNR. In Sec. \ref{SEC_ASER}, we study the ASER performance of our system in the case of imperfect channel estimation. We develop upper and lower bound analysis for the general relay case to formulate the relay selection scheme. We then derive a closed-form formula for the ASER performance with imperfect channel estimation. We also derive closed-form formula for the ASER performance of our system in the high SNR regime as introduced in \cite{Giannakis2005,Giannakis2003}. We also study the outage probability and average capacity per bandwidth in Sec.~\ref{SEC_outage} and Sec.~\ref{SEC_Capacity}, respectively. Diversity analysis is also provided in \ref{SEC_Diversity}. Simulation results are presented in Sec. \ref{SEC_SIMULATION}, followed by conclusions in Sec. \ref{SEC_CONCLUSION}.

\textit{Notation}: ${\rm E}\{\cdot\}$, $\{\cdot\}^*$ and $\hat{x}$, denote the expected value, the complex conjugate and the estimate of the variable $x$, respectively. $\gamma_{_{i,lb}}$ and $\gamma_{_{i,ub}}$ stand for the $i^{th}$ lower bound and upper bound $\gamma$, respectively. $\gamma_{i^\star}$ indicates that the relay $i$ with SNR $\gamma$ is selected. {$f_x(x)$ and $F_x(x)$ denote the  probability density function (PDF) and the cumulative density function (CDF), respectively.}
\section{system model}\label{sec system model}
In this section we consider a system in which a source node ${\it S}$ transmits information to a destination node ${\it D}$ with the help of the best relay node ${\it R_{i^{{\tiny\star}}}}$, which is selected under a defined criterion from {\it M} available relays. The transmissions are orthogonal, either through time or frequency division.

In the first data sharing time slot, the source node communicates with the destination as well as the relay nodes. In this phase, the signals received by the destination and each relay are
\begin{eqnarray}
y_{_{sd}}\!\!\!&=&\!\!\!\sqrt{ {\mathcal{E}_{_{sd}}}}h_{_{sd}}x+n_{_{sd}},\label{EQ_y_sd}\\
y_{_{si}}\!\!\!&=&\!\!\!\sqrt{{\mathcal{E}_{_{si}}}}h_{_{si}}x+n_{_{si}},\label{EQ_y_si}
\end{eqnarray}
where $x$, $y_{_{sd}}$, and $y_{_{si}}$ denote the transmitted signal with unit energy and the signals received at the destination and the $i^{th}$ relay node, respectively. $h_{_{si}}$ and $h_{_{sd}}$ are the channel coefficients of the source-relay and the source-destination channels, which include the effect of fading. ${\mathcal{E}_{_{sd}}}$ and ${\mathcal{E}_{_{si}}}$ represent the average
signal energies received at destination and the relay $i$ terminals, respectively, taking into
account the path loss and shadowing effects. $n_{_{sd}}$ and $n_{_{si}}$ are additive white Gaussian noises (AWGN) in the corresponding channels with the same variance ${\it N_0}$, i.e., $n_{_{sd}}, n_{_{si}}\sim\mathcal{C}\mathcal{N}(0,{\it N}_0)$.

In the next time slot each relay node normalizes its received signal and retransmits it to the destination. For the $i^{th}$ relay, the normalization factor is $\NOP$ and the signal transmitted from the this relay is \cite{Adve2007}
\begin{equation}\label{EQ_XI}
x_i=\frac{\sqrt{{\it {\mathcal{E}_{_{si}}}}}h_{_{si}}x+n_{_{si}}}{\NOP}.
\end{equation}
Since the relays practically do not have a perfect knowledge about the channel $h_{_{si}}$, $i=1, \ldots, M$, the power normalization procedure in each relay is performed based on the estimate of the corresponding channel. We assume in this paper that ${\rm E}\{|x|^2\}=1$.
Based on (\ref{EQ_XI}), the signal received by the destination from the $i^{th}$ relay node is \cite{Adve2007}
\begin{eqnarray}\label{EQ_Y_SEL}
y_{_{id}}\!\!\!&=&\!\!\!\sqrt{{\it {\mathcal{E}_{_{id}}}}}h_{_{id}}x_i+n_{_{id}},\nonumber\\
\!\!\!&=&\!\!\!\frac{\sqrt{\it {\mathcal{E}_{_{si}}} {\mathcal{E}_{_{id}}}}}{\NOP}h_{_{si}}h_{_{id}}x+\tilde{n}_{_{id}},\\
\!\!\!&=&\!\!\!\alpha_ih_{_{si}}h_{_{id}}x+\tilde{n}_{_{id}}.
\end{eqnarray}
where $h_{_{id}}$ is the channel gain from this node to the destination. ${\mathcal{E}_{_{id}}}$ is the average signal energy received at the destination via the $i^{th}$ relay in its corresponding time slot which accounts for the shadowing and path loss effects and $\alpha_i=\frac{\sqrt{\it {\mathcal{E}_{_{si}}} {\mathcal{E}_{_{id}}}}}{\NOP}$. $n_{_{id}}\sim\mathcal{CN}(0,{\it N_0})$ denotes the AWGN of the relay destination channel. $\tilde{n}_{_{id}}$ is the equivalent noise term in $y_{_{id}}$. {It can easily be shown that, conditioned on the channel realizations $\tilde{n}_{_{id}}\sim\mathcal{CN}(0,\omega_i^2{\it N_0})$, where}
\begin{equation}\label{EQ_OMEGA1}
\omega_i^2=1+\frac{{ {\mathcal{E}_{_{id}}}}|h_{_{id}}|^2}{{\mathcal{E}_{_{si}}}|\hat{h}_{_{si}}|^2+N_0}.
\end{equation}

 Supposing only the $i^{th}$ relay participates in cooperation we look for the condition under which we can select the best relay, $R_{i^{{\tiny\star}}}$, by searching $R_i$s for $i=1, \ldots, M$.

 Since each relay only amplifies the signal from the source, the destination is the only place where an estimate of the information symbol $x$ is computed. In order to achieve maximum likelihood performance, the signals from both diversity branches (direct $S\rightarrow D$ branch and the branch via the $i^{th}$ relay) are combined using a maximum ratio combiner (MRC). Decoding of $x$ is delayed until the relayed signal containing the information symbol $x$ is received at the destination. Since the noise power is not the same on the two sub channels, both diversity branches must be weighted by their respective complex fading gain over total noise power on that particular branch before the combiner. Thus we obtain the estimated information symbol as
\begin{eqnarray}\label{EQ_MRC}
\hat x\!\!\!&=&\!\!\!\frac{\hat{h}_{_{sd}}^*\sqrt{{\mathcal{E}_{_{sd}}}}}{{N_0}}y_{_{sd}}+\frac{\alpha_i^*\hat{h}_{_{si}}^*\hat{h}_{_{id}}^*}{\hat{\omega}^2_iN_0}y_{_{id}},
\end{eqnarray}
where
\begin{equation}\label{EQ_OMEGA}
\hat{\omega}_i=\sqrt{\frac{{ {\mathcal{E}_{_{si}}}}|\hat{h}_{_{si}}|^2+{{\mathcal{E}_{_{id}}}}|\hat{h}_{_{id}}|^2+N_0}{{ {\mathcal{E}_{_{si}}}}|\hat{h}_{_{si}}|^2+N_0}},
\end{equation}
since the destination only uses the estimate of the relay-destination channel.

\section{Modeling of channel estimation error}\label{Sec_Error model}
Let the true channel gain and its estimate be $h\sim\mathcal{CN}(0,\sigma_h^2)$ and $\hat h\sim\mathcal{CN}(0,\sigma^2_{\hat {h}})$, respectively. It is shown in \cite{Leung2003,Ahmed2006} that it is possible to consider the channel estimate as
\begin{equation}\label{EQ_Erro}
\hat{h}= h+e,
\end{equation}
where $e\sim\mathcal{CN}(0,\it \sigma_e^2)$ is a zero mean complex Gaussian noise. As $h$ and $\hat h$, are jointly Gaussian, the conditional probability density function (PDF), $f(h|\hat h)$, would also be Gaussian with the mean \cite{Kay1998}
\begin{equation}\label{EQ_meanCOnditional}
{\rm E}\{h|\hat h\}=\frac{\sigma_h^2}{\sigma_{\hat h}^2}\hat h,
\end{equation}
and variance of
\begin{equation}\label{EQ_variance}
{\rm var}\{h|\hat h\}=(1-\frac{\sigma_h^2}{\sigma_{\hat h}^2})\sigma_h^2.
\end{equation}
By (\ref{EQ_meanCOnditional}) and (\ref{EQ_variance}) and assuming a least squares (LS) estimator, we can consider the channel as
\begin{equation}\label{EQ_rho}
h=\rho\hat h+d
\end{equation}
{where $\rho =(\sigma_{h}^2/\sigma_{\hat h}^2)=\rho_{h\hat{h}}(\sigma_h/\sigma_{\hat{h}})$ $\rho_{h\hat{h}}$ is the correlation coefficient between $h$ and $\hat{h}$ and $d\sim\mathcal{CN}(0,\sigma_D^2)$. In this paper, $\sigma^2_{D,sd}, \sigma^2_{D,si}, \sigma^2_{D,id}, \rho_{sd}, \rho_{si}$ and $\rho_{id}$ are the channel estimation error variances and their corresponding $\rho$ factors of the $S-D$, $S-R_i$, and $R_i-D$ links, respectively.}

According to (\ref{EQ_variance})
\begin{eqnarray}\label{EQ_rho1}
\sigma_D^2&=&(1-\rho)\sigma_h^2\no\\
&=&(\rho-\rho^2)\sigma_{\hat{h}}^2.
\end{eqnarray}
Using (\ref{EQ_rho}), we can rewrite the first term of (\ref{EQ_MRC}) as
\begin{eqnarray}\label{EQ_MRC_SD}
D_{_{sd}}\!\!\!&=&\!\!\!\frac{\hat{h}_{_{sd}}^*\sqrt{{\mathcal{E}_{_{sd}}}}}{ N_0}\left(\!\!\!\begin{array}{c}\sqrt{ {\mathcal{E}_{_{sd}}}}(\rho_{_{sd}}\hat{h}_{_{sd}}+d_{_{sd}})x+n_{_{sd}}\end{array}\!\!\!\right)\no\\
\!\!\!&=&\!\!\!\frac{\rho_{_{sd}}{{\mathcal{E}_{_{sd}}}}|\hat{h}_{_{sd}}|^2}{\it N_0}x+\frac{\sqrt{{\mathcal{E}_{_{sd}}}}\hat h_{_{sd}}^*}{\it N_0}(\sqrt{{\mathcal{E}_{_{sd}}}} d_{_{sd}}x+n_{_{sd}}).
\end{eqnarray}
Here, we have inserted (\ref{EQ_rho}) in (\ref{EQ_y_sd}) then the resultant expression is substituted into (\ref{EQ_MRC}).
Following \cite{Han2009}, the received direct path signal after the MRC can be decomposed into the message part given as $\frac{\rho_{_{sd}}{\it {\mathcal{E}_{_{sd}}}}|\hat{h}_{_{sd}}|^2}{\it N_0}$ and the noise part which is given as $\frac{\sqrt{ {\mathcal{E}_{_{sd}}}}\hat h_{_{sd}}^*}{\it N_0}(\sqrt{{\mathcal{E}_{_{sd}}}} d_{_{sd}}+n_{_{sd}})$.
Finally, due to the independence of the $\hat{h}_{_{sd}}$, $d_{_{sd}}$ and $n_{_{sd}}$, the effective SNR due to the direct path is given by
\begin{eqnarray}\label{EQ_EFFSNR_SD}
\gamma_{_{sd}}^{\rm eff}&=&\frac{\rE_x\{|{\rm E}\{D_{_{sd}}\}|^2\}}{{\rm var}\{D_{_{sd}}\}}\no\\
&=&\frac{\frac{{\mathcal{E}_{_{sd}}}}{N_0}\rho_{_{sd}}^2|\hat{h}_{_{sd}}|^2}{1+\frac{{\mathcal{E}_{_{sd}}}}{N_0}\sigma^2_{D,sd}}=\frac{\rho_{_{sd}}^2\hat{\gamma}_{_{sd}}}{1+\epsilon_{_{sd}}},
\end{eqnarray}
 where $\hat{\gamma}_{_{sd}}$ is the estimated received SNR due to the direct path and $\epsilon_{_{sd}}=\frac{{\mathcal{E}_{_{sd}}}}{N_0}\sigma^2_{D,sd}$.

 Following the same approach, we can write the second term of (\ref{EQ_MRC}) as
\begin{eqnarray}\label{EQ_MRC_ID}
D_{_{id}}&&\!\!\!\!\!\!\!\!\!=\frac{\alpha_i\hat{h}_{_{si}}^*\hat{h}_{_{id}}^*}{\hat{\omega}^2_iN_0}\left[\!\!\!\begin{array}{c}\alpha_i(\rho_{_{si}}\hat{h}_{_{si}}+d_{_{si}})(\rho_{_{id}} \hat{h}_{_{id}}+d_{_{id}})x+\tilde{n}_{_{id}}\end{array}\!\!\!\right].
\end{eqnarray}
We can extract the message part, error part due to channel estimation error, and the noise part of the received signal as\footnote{{Conditioned on $h_{_{si}}$ and  $h_{_{id}}$ the distribution of (\ref{EQ_D}) is approximated as Gaussian. The validity of this approximation has been confirmed by simulation. It should be further noted that  the same conclusion has been reported by \cite{Han2009} and the references therein.}}
\begin{eqnarray}\label{EQ_message part}
&&\mathcal{M}=\frac{\alpha_i^2}{\hat{\omega}_i^2N_0}\rho_{_{si}}\rho_{_{id}}|\hat{h}_{_{si}}|^2|\hat{h}_{_{id}}|^2x,\\
&&\mathcal{D}=\frac{\alpha_i^2}{\hat{\omega}_i^2N_0}[\rho_{_{si}}|h_{_{si}}|^2h_{_{id}}^*d_{_{id}}+\rho_{_{id}}|h_{_{id}}|^2h_{_{si}}^*d_{_{si}}\label{EQ_D}\no\\
&&+h^*_{_{si}}h^*_{_{id}}d_{_{si}}d_{_{id}}]x,\\
&&\mathcal{N}=\frac{\alpha_i\hat{h}_{_{si}}^*\hat{h}_{_{id}}^*}{\hat{\omega}^2_iN_0}\tilde{n}_{_{id}},\label{EQ_NN}
\end{eqnarray}
respectively. To obtain the effective SNR expression, we need to find the ratio of the signal power to the overall noise power. {Alternatively, the received SNR can be calculated as  $\frac{\rE_x\{|{\rm E}\{D_{id}\}|^2\}}{{\rm var}\{D_{id}\}}$\cite{Chen2005}.}
Since $h_{_{si}}$, $h_{_{id}}$, $d_{_{si}}$ and  $d_{_{id}}$ are zero mean independent processes, we can write the effective SNR of the selected path as
\vspace{1mm}
\begin{eqnarray}\label{EQ_gamma_Eff11}
\gamma_{_i}^{\rm eff}=\frac{{\mathcal{E}_{_{si}}}\rho_{_{si}}^2|\hat{h}_{_{si}}|^2{\mathcal{E}_{_{id}}}\rho_{_{id}}^2|\hat{h}_{_{id}}|^2}{N_0^2(\frac{{\mathcal{E}_{_{si}}}}{N_0}|\hat{h}_{_{si}}|^2(1+\rho_{_{si}}^2\frac{{\mathcal{E}_{_{id}}}}{N_0}\sigma^2_{D,id})+\frac{{\mathcal{E}_{_{id}}}}{N_0}|\hat{h}_{_{id}}|^2(1+\rho_{_{id}}^2\frac{{\mathcal{E}_{_{si}}}}{N_0}\sigma^2_{D,si})+\frac{{\mathcal{E}_{_{si}}}}{N_0}\sigma^2_{D,si}\frac{{\mathcal{E}_{_{id}}}}{N_0}\sigma^2_{D,id}+1)}.
\end{eqnarray}

\vspace{5mm}

With a simple manipulation, we may write (\ref{EQ_gamma_Eff11}) as follows
\begin{eqnarray}\label{EQ_gamma_Eff2}
\gamma_{_i}^{\rm eff}=\frac{\rho_{_{si}}^2\rho_{_{id}}^2\hat{\gamma}_{_{si}}\hat{\gamma}_{_{id}}}{\hat{\gamma}_{_{si}}\lambda_{_{si}}+\hat{\gamma}_{_{id}}\lambda_{_{id}}+\epsilon_{_{si}}\epsilon_{_{id}}+1},
\end{eqnarray}
where $\lambda_{_{si}}=1+\rho_{_{si}}^2\epsilon_{_{id}}$, $\lambda_{_{id}}=1+\rho_{_{id}}^2\epsilon_{_{si}}$, $\hat{\gamma}_{_{si}}=\frac{{\mathcal{E}_{_{si}}}|\hat{h}_{_{si}}|^2}{N_0}$, $\hat{\gamma}_{_{id}}=\frac{{\mathcal{E}_{_{id}}}|\hat{h}_{_{id}}|^2}{N_0}$, $\epsilon_{_{si}}=\frac{{\mathcal{E}_{_{si}}}\sigma^2_{D,si}}{N_0}$, and $\epsilon_{_{id}}=\frac{{\mathcal{E}_{_{id}}}\sigma^2_{D,id}}{N_0}$. Following \cite{Han2009}, $1+\epsilon_{_{si}}\epsilon_{_{id}}$
can be ignored \footnote{{For a fixed SNR, i.e., $\frac{\cE}{N_0}=5$dB, we have noticed that the mean square error MSE between $\gamma_i^{\rm eff}$ in (\ref{EQ_gamma_Eff2}) and (\ref{EQ_gamma_Eff3}) is $10^{-10}$ percent of $\gamma_i^{\rm eff^2}$ for $\hat{\gamma}_{_{si}}=\hat{\gamma}_{_{id}}=30$dB.}}; hence, (\ref{EQ_gamma_Eff3}) can be approximated as \cite{Anghel2004,Han2009}

\begin{eqnarray}\label{EQ_gamma_Eff3}
\gamma_{_i}^{\rm eff}=\frac{\rho_{_{si}}^2\rho_{_{id}}^2\hat{\gamma}_{_{si}}\hat{\gamma}_{_{id}}}{\hat{\gamma}_{_{si}}\lambda_{_{si}}+\hat{\gamma}_{_{id}}\lambda_{_{id}}}.
\end{eqnarray}

{\section{selection Strategy}}
In the selection based amplify-and-forward scheme studied in \cite{Giannakis2005,Anghel2004,Adve2007,Adve2006,Beres2008,VTC2010,QBSC2010} the best relay selected by the destination terminal is the one that leads to the maximum total received SNR, $\gamma_{_r}$, which is given by
\begin{eqnarray}\label{EQ_GAmma_TOTAL}
\gamma_{_r}\!\!\!&=&\!\!\!\gamma_{_{sd}}^{\rm eff}+\gamma_{i^{{\tiny\star}}}^{\rm eff},%
\end{eqnarray}
where
\begin{equation}
\gamma_{i^{{\tiny\star}}}^{\rm eff}=\arg ~\max_i\{\gamma_{i}^{\rm eff}\},
\end{equation}
for $i=1\ldots M$.{ $\gamma_r$ , which is the total received SNR at the destination, is the addition of the SNR received from the direct path,
$\gamma_{_{sd}}^{\rm eff}$, and the SNR received via the selected relay path $\gamma_{i^\star}^{\rm eff}$. The selection steps are
mainly as follows. The source terminal sends a ready to send (RTS) packet to all the relays and the destination. As the relays receive this packet they estimate
 their corresponding source-to-relay channel $h_{si}$. They divide their signal by the factor $\sqrt{\cE_{si}|\hat{h}_{si}|^2+N_0}$ and wait for the clear to send
 (CTS) packet from the receiver side. As each relay receives the CTS overhead, it estimates its corresponding relay-to-destination $h_{id}$ channel and will start
 a timer which is given by
 $$\tau_i=\frac{1}{\gamma_i^{\rm eff}},$$
 where $\gamma_i^{\rm eff}$ is given by (\ref{EQ_gamma_Eff2}). Each timer $\tau_i$ counts down to zero right after
 receiving the CTS packet. The first timer that reaches zero sends a flag to all the relays and the destination, introducing itself as the best relay. Then all the
 relays would remain silent, and the selected relay forwards its signal to the destination, with an overhead package containing information about its source-to-relay
  link ${h}_{si}$ and the relay-to-destination channel ${h}_{id}$. It is also possible for the relays to use a separate $\log_2(|\hat{h}_{si}|^2)+\log_2(|\hat{h}_{id}|^2)$
   bit feedback link to send the CSI to the destination. Further details on the selection procedures can be found in
\cite{Bletsas2006,Jaffarkhani2009}.}

\section{ASER analysis}\label{SEC_ASER}

In this section, we derive an ASER expression for the relay selection scheme. We first introduce the following theorems \cite{Papoulis2002}
\begin{itemize}
\item {\emph{Theorem 1}}:  If $X$ and $Y$ are two random variables with $Y=mX$, then
\begin{equation}
f_Y(\gamma)=\frac1{|m|}f_X(\frac{\gamma}{m}).\no
\end{equation}

\item {\emph{Theorem 2}}:  For two independent random variables $X$ and $Y$ if $Z=\min(X,Y)$, then
\begin{equation}
f_Z(\gamma)=f_X(\gamma)+f_Y(\gamma)-f_X(\gamma)F_Y(\gamma)-f_Y(\gamma)F_X(\gamma).\no
\end{equation}
\item {\emph{Theorem 3}}: Let $Y=\max_i(X_i)$, $i=1, \cdots M$, where $X_i$s are independent random variables, then
\begin{equation}
F_Y(\gamma)=\prod_{i=1}^MF_{X_i}(\gamma).\no
\end{equation}
\end{itemize}

It is shown in \cite{Giannakis2005,Anghel2004,Adve2006,Ikkic2009,Adve2006,Ikki2008,Ikki2009,Ding} that the symbol error rate (SER) in a cooperative scenario is given by $Q(\sqrt{k\gamma_{_r}})$,
where
$$
Q(x)=\frac1{\sqrt{2\pi}}\int_x^{\infty}\exp(\frac{-t^2}{2})~dt,
$$
and $k$ depends on the modulation scheme.\footnote{For instance for BPSK modulation $k=2$. The analysis can be generalized to QPSK or QPAM modulation as well\cite{Giannakis2005}.}

To find the average symbol error rate, we need to integrate $Q(\sqrt{k\gamma_{_r}})$ over the probability density function (PDF) of $\gamma_{_r}$, i.e.,
\begin{equation}\label{EQ_pe}
\bar{P}_e=\int_0^{\infty}Q(\sqrt{k\gamma_{_r}})~f_{\gamma_{_r}}(\gamma_{_r})~d\gamma_{_r}.
\end{equation}

Finding the exact expression for $f_{\gamma_{_r}}$ can be quite cumbersome, therefore, in the following and to simplify analysis, we develop lower and upper bounds on $\gamma_{_r}$.
It is straightforward to show that for two arbitrary independent random variables, the following inequality holds \cite{Anghel2004}
\begin{equation}\label{EQ_xybound}
\frac1{2}\min{(X,Y)}\leq\frac{XY}{X+Y}\leq\min{(X,Y)}.
\end{equation}
Applying (\ref{EQ_xybound}) to $\gamma_{_i}^{\rm eff}$ in (\ref{EQ_gamma_Eff3}) yields
\begin{equation}\label{EQ_gamma_i_bound}
\gamma_{_{i,lb}}\leq\gamma_{_i}^{\rm eff}\leq\gamma_{_{i,ub}},
\end{equation}
where
\begin{eqnarray}
\gamma_{_{i,ub}}\!\!\!&=&\!\!\!\frac{\rho^2_{_{si}}\rho^2_{_{id}}}{\lambda_{_{si}}\lambda_{_{id}}}\min{(\hat{\gamma}_{_{si}}\lambda_{_{si}},\hat{\gamma}_{_{id}}\lambda_{_{id}})},\label{EQ_uub}\\
\gamma_{_{i,lb}}\!\!\!&=&\!\!\!\frac12\frac{\rho^2_{_{si}}\rho^2_{_{id}}}{\lambda_{_{si}}\lambda_{_{id}}}\min{(\hat{\gamma}_{_{si}}\lambda_{_{si}},\hat{\gamma}_{_{id}}\lambda_{_{id}})},\label{EQ_llb}
\end{eqnarray}
Finally, $\gamma_{_r}$, can be bounded as
\begin{equation}\label{EQ_ban}
\gamma_{_{lb}}\leq\gamma_{_r}\leq\gamma_{_{ub}},
\end{equation}
where $\gamma_{_{lb}}$ and $\gamma_{_{ub}}$ are the lower bound and the upper bound SNR values defined as
\begin{eqnarray}\label{EQ_LBUB}
\gamma_{_{ub}}\!\!\!&=&\!\!\!\frac{\rho_{_{sd}}^2\hat{\gamma}_{_{sd}}}{1+\epsilon_{_{sd}}}+\gamma_{_{i^{{\tiny\star}}\!\!,ub}},\label{EQ_LBUB1}
\\\gamma_{_{lb}}\!\!\!&=&\!\!\!\frac{\rho_{_{sd}}^2\hat{\gamma}_{_{sd}}}{1+\epsilon_{_{sd}}}+\gamma_{_{i^{{\tiny\star}}\!\!,lb}},\label{EQ_LBUB2}
\end{eqnarray}
respectively, where
\begin{eqnarray}
\gamma_{_{i^{{\tiny\star}}\!\!,ub}}&\stackrel{\Delta}{=}&\max_i{\{\gamma_{_{i,ub}}\}},\\
\gamma_{_{i^{{\tiny\star}}\!\!,lb}}&\stackrel{\Delta}{=}&\max_i{\{\gamma_{_{i,lb}}\}}.
\end{eqnarray}

Using (\ref{EQ_ban}), (\ref{EQ_LBUB1}) and (\ref{EQ_LBUB2}), (\ref{EQ_pe}) can be bounded as
\begin{equation}\label{EQ_P_UB}
\bar{P}_{_{lb}}\leq\bar{P}_e\leq\bar{P}_{_{ub}},
\end{equation}
where $\bar{P}_{_{lb}}$ and $\bar{P}_{_{ub}}$, are the lower and the upper bounds for ASER's. In this paper, we limit our discussion to the lower bound  analysis. The upper bound analysis can be obtained similarly. Since the ASER is a decreasing function of SNR, a lower bound for $\int_0^\infty Q(\sqrt{k\gamma_{_r}})f_{\gamma_{_r}}d\gamma_{_r}$ would simply be $\int_0^{\infty} Q(\sqrt{k\gamma_{_{ub}}})f_{\gamma_{_{ub}}}d\gamma_{_{ub}}$.

Since our final goal is to obtain the PDF of $\gamma_{_{ub}}$ we need to find the PDF of $\gamma_{_{i,ub}}$, for $i=1, \ldots M$, first. Note that the PDF of $\hat{\gamma}_{_{si}}$ can be expressed in terms of the average SNR $\bar{\gamma}_{_{si}}=\frac{{\mathcal{E}_{_{si}}}{\rm E}\{|\hat{h}_{_{si}}|^2\}}{N_0}$ as $f_{\hat{\gamma}_{_{si}}}(\gamma)=\frac1{\bar{\gamma}_{_{si}}}\exp{(\frac{-\gamma}{\bar{\gamma}_{_{si}}})}$.
Similar expressions can be found for the PDF of $\hat{\gamma}_{_{sd}}$ and $\hat{\gamma}_{_{id}}$ as well.
Thus, using \emph{Theorem 1}, the PDF of ${\gamma_{_{i,ub}}}$ can be written as \footnote{Here we have to note that ${\gamma_{_{i,ub}}}$s are not i.i.d random variables, since $\beta_i$s are not analogous for different relay paths.}
\begin{equation}\label{EQ_BETTA_i}
f_{\gamma_{_{i,ub}}}(\gamma)=\frac1{\beta_i}\exp(\frac{-\gamma}{\beta_i}),
\end{equation}
where $\beta_i=\frac{(\bar{\gamma}_{_{si}}\bar{\gamma}_{_{id}})\rho_{_{si}}^2\rho_{_{id}}^2}{\lambda_{_{si}}\bar{\gamma}_{_{si}}+\lambda_{_{id}}\bar{\gamma}_{_{id}}}$. Using \emph{Theorem 2} and noting that $\gamma_{_{i^{{\tiny\star}}\!\!,ub}}=\max_i{(\gamma_{_{i,ub}})}$, we obtain
\begin{eqnarray}
f_{\gamma_{_{i^{{\tiny\star}}\!\!,ub}}}(\gamma)\!\!\!&=&\!\!\!\frac{\partial F_{\gamma_{_{i^{{\tiny\star}}\!\!,ub}}(\gamma)}}{\partial\gamma}\\
&=&\!\!\!\ \frac{\partial \prod_{i=1}^M F_{\gamma_{_{i,ub}}}(\gamma)}{\partial\gamma}\\
&=&\!\!\!\ \sum_{l=1}^M f_{\gamma_{_{l,ub}}(\gamma)\!\!\!\!}\prod_{{\tiny\begin{array}{c}i=1\\i\neq l\end{array}}}^M\!\!\!\! F_{\gamma_{_{i,ub}}}(\gamma).\label{EQ_theta_ub}
\end{eqnarray}

The closed-form formulation for $f_{\gamma_{_{i^{{\tiny\star}}\!\!,ub}}}(\gamma)$ would then be (see Appendix~I)

\begin{eqnarray}
f_{\gamma_{_{ub}}}(\gamma)=\sum_{p=1}^M\sum_{m=1}^{{\tiny \left(\!\!\!\!\begin{array}{c}M\\p\end{array}\!\!\!\!\right)}}(-1)^{(p+1)}\frac{[\Psi^{p}\bb]_m}{\bar{\mu}[\Psi^{p}\bb]_m-1}\left[\begin{array}{c}\!\!\!\exp(\frac{-\gamma}{\bar{\mu}})-\exp(-\gamma[ \Psi^{p}\bb]_m)\end{array}\!\!\!\right],\label{EQ_MAM}
\end{eqnarray}
{where $\bar{\mu}=\frac{\rho_{_{sd}}^2\bar{\gamma}_{_{sd}}}{1+\epsilon_{_{sd}}}$ and $\Psi^{p}$ is a binary permutation matrix with dimension of ${{\tiny\left(\!\!\!\begin{array}{c}M\\p\end{array}\!\!\!\right)\times M}}$ which is defined in the Appendix~I and
\begin{equation}\label{EQ_b}
\bb\stackrel{\Delta}{=}[\frac1{\beta_1}~\frac1{\beta_2}~\ldots~\frac1{\beta_M}]^T,
\end{equation}
}
\begin{equation}
Q(\gamma)=\frac12-\frac12 {\rm Erf}(\frac{\gamma}{\sqrt{2}}).
\end{equation}
Having noted that
\begin{equation}\label{EQ_Plb}
\bar{P}_{_{lb}}=\int_0^\infty \left[\!\!\!\begin{array}{c}\frac12-\frac12{\rm Erf}(\sqrt{\frac{k}2\gamma_{_{ub}}})\end{array}\!\!\!\right]f_{\gamma_{_{ub}}}(\gamma_{_{ub}})d_{\gamma_{_{ub}}}
\end{equation}
by substituting (\ref{EQ_MAM}) into (\ref{EQ_Plb}) and setting $\gamma_{_{ub}}=\gamma^2$,  we can write $\bar{P}_{_{lb}}$ as
\begin{eqnarray}¸\label{EQ_FINN}
\bar{P}_{_{lb}}=\frac12-\frac12\sum_{p=1}^M\sum_{m=1}^{{\tiny \left(\!\!\!\!\begin{array}{c}M\\p\end{array}\!\!\!\!\right)}}(-1)^{(p+1)}\frac{[\Psi^{p}\bb]_m}{\bar{\mu}[\Psi^{p}\bb]_m-1}\left[\begin{array}{c}\bar{\mu}\sqrt{\frac{k/2}{\bar{\mu}^{-1}+k/2}}-[\Psi^{p}\bb]_m^{-1}\sqrt{\frac{k/2}{[\Psi^{p}\bb]_m+k/2}}\end{array}\right].
\end{eqnarray}
Here, we have used the fact \cite{Gradshteyn}
\begin{equation}
\int_0^\infty \gamma~ {\rm Erf}(a\gamma) \exp{(-b^2\gamma^2)}d\gamma=\frac{a}{2b^2\sqrt{b^2+a^2}} ~~~~b\neq0.
\end{equation}
Note that the upper bound analysis can be derived similarly.

To gain further insight into the performance of the selection scheme under consideration, we resort to the high SNR analysis. {Defining $\hat{\gamma}_{_{ub}}=\gamma_{_{ub}}/\bar{\gamma}_{_{ub}}$, where $\bar{\gamma}_{_{ub}}$ is the average SNR, and following \cite{Giannakis2003,Giannakis2005}, we can approximate the asymptotic behavior of $f_{\hat{\gamma}_{_{ub}}}$ in the high SNR regime with its Mclaurin series expansion, i.e.,}

{
\begin{equation}\label{EQ_mclaurin}
f_{\hat{\gamma}_{_{ub}}}=a\hat{\gamma}_{_{ub}}^n+o(\hat{\gamma}_{_{ub}}),
\end{equation}
where $a=\frac1{n!}\frac{\partial^nf_{\hat{\gamma}_{_{ub}}}}{\partial\hat{\gamma}_{_{ub}}^n}(0)$, if the derivatives of $f_{\hat{\gamma}_{_{ub}}}$ up to order $n-1$ are null and
$$
o(\hat{\gamma}_{_{ub}})=\sum_{t=n+1}^{\infty}\frac1{t!}\frac{\partial^{t}f_{\hat{\gamma}_{_{ub}}}}{\partial\hat{\gamma}_{_{ub}}^{t}}(0)\hat{\gamma}_{_{ub}}^{t}.
$$
It can be easily shown that by replacing (\ref{EQ_mclaurin}) into (\ref{EQ_Plb}), the asymptotic behavior of ASER  $(\bar{\gamma}_{_{ub}}\rightarrow\infty)$ is given by \cite{Giannakis2005}
\begin{equation}\label{EQ_ass_P_E}
\bar{P}_{_{lb}}\rightarrow\frac{\prod_{m=1}^{n+1}(2m-1)}{2(n+1)k^{(n+1)}}\frac1{n!}\frac{\partial^nf_{\gamma_{_{ub}}}}{\partial\gamma_{_{ub}}^n}(0).
\end{equation}
where we have used the fact that $f_{\gamma_{_{ub}}}(\gamma_{_{ub}})=\frac{1}{\bar{\gamma}_{_{ub}}}f_{\hat{\gamma}_{_{ub}}}(\frac{\gamma_{_{ub}}}{\bar{\gamma}_{_{ub}}})$.}

{Similarly, the asymptotic upper bound ASER can be found as
\begin{equation}\label{EQ_ass_P_El}
\bar{P}_{_{ub}}\rightarrow\frac{\prod_{m=1}^{n+1}(2m-1)}{2(n+1)k^{(n+1)}}\frac1{n!}\frac{\partial^nf_{\gamma_{_{lb}}}}{\partial\gamma_{_{lb}}^n}(0).
\end{equation}
}

\emph{Theorem 4:} For all $n<M$, the values of $\frac{\partial^nf_{\gamma_{_{ub}}}}{\partial\gamma_{_{ub}}^n}(0)$ and $\frac{\partial^nf_{\gamma_{_{lb}}}}{\partial\gamma_{_{lb}}^n}(0)$ are zero and for $n=M$
\begin{eqnarray}
\bar{P}_{_{lb}}\rightarrow\frac{\prod_{m=1}^{M+1}(2m-1)}{2(M+1)k^{(M+1)}}\frac1{\mu}\prod_{i=1}^{M}\frac1{\beta_i}\label{EQ_THEORY4},\\
{\bar{P}_{_{ub}}\rightarrow\frac{\prod_{m=1}^{M+1}(2m-1)}{(M+1)k^{(M+1)}}\frac1{\mu}\prod_{i=1}^{M}\frac1{\beta_i},}\label{EQ_THEORY42}
\end{eqnarray}
{where $\mu=\frac{\rho_{_{sd}}^2\hat{\gamma}_{_{sd}}}{1+\epsilon_{_{sd}}}$.}

\emph{Proof of Theorem 4:} (see Appendix~II)

\section{Outage probability Analysis}\label{SEC_outage}

The outage probability $P_{out}$, {which is a valid measure of performance in slowly fading channels}, is defined as the probability that the source-destination mutual information $\mathcal{I}_{_{sd}}$ falls below the transmission rate $R$. Defining
\begin{equation}
\mathcal{I}_{_{sd}}\stackrel{\Delta}{=}\frac12\log_2(1+\gamma_{_{sd}}^{\rm eff}+\gamma_{i^{{\tiny\star}}}^{\rm eff})\leq R,
\end{equation}
the outage probability is given by
\begin{eqnarray}
P_{out}&=&P\left(\begin{array}{c}\!\!\! \frac12\log_2\left(\begin{array}{c}\!\!\!1+\gamma_{_r}\!\!\!\end{array}\right)\leq R\!\!\! \end{array}\right)\no\\
&=&P\left(\begin{array}{c}\!\!\! \gamma_{_r}\leq 2^{2R}-1\!\!\!\end{array}\right).
\end{eqnarray}
Substituting the introduced upper bound for $\gamma_{_r}$ leads to adopt an upper bound on the outage probability, i.e.,
\begin{eqnarray}
P_{out}\leq P_{out}^{ub}&=&P\left(\begin{array}{c}\!\!\! \gamma_{_{ub}}\leq 2^{2R}-1\!\!\!\end{array}\right)\no\\
&=&F_{\gamma_{_{ub}}}(2^{2R}-1).
\end{eqnarray}
We can simply find $F_{\gamma_{_{ub}}}$ by integrating (\ref{EQ_MAM}) with respect to $\gamma$ the result of which is given by
\begin{eqnarray}¸\label{EQ_CDF}
F_{\gamma_{_{ub}}}(\gamma)\!\!\!&=&\!\!\!\sum_{p=1}^M\sum_{m=1}^{{\tiny \left(\!\!\!\!\begin{array}{c}M\\p\end{array}\!\!\!\!\right)}}(-1)^{(p+1)}\left[\begin{array}{c}\!\!\!1+\frac{1}{\bar{\mu}[\Psi^{p}\bb]_m-1}\exp(-\gamma[\Psi^{p}\bb]_m)-\frac{\bar{\mu}[\Psi^{p}\bb]_m}{\bar{\mu}[\Psi^{p}\bb]_m-1}\exp(\frac{-\gamma}{\bar{\mu}})\end{array}\!\!\!\right].\no\\
\end{eqnarray}
We can apply the same procedure for the lower bound analysis.

\section{Average Capacity}\label{SEC_Capacity}
{ The maximum achievable transmission rate under which the occurred errors can be recovered is called the system ergodic capacity. The average capacity per time slot for S-AF scheme for the perfect CSI case is defined as \cite{Costa2007}
\newcounter{mytempeqncnt121}
\normalsize \setcounter{mytempeqncnt121}{\value{equation}}
\setcounter{equation}{61}
\begin{equation}\label{EQ_Capacity_perf}
\bar{C}=\frac{W}2\int_0^{\infty}\log_2(1+\gamma_{_{t}})f_{\gamma_{_{t}}}(\gamma_{_{t}})d\gamma_{_t}
\end{equation}
in bits per time slot, where $W$ is the transmitted signal bandwidth, and $\gamma_{_t}$ is the total received SNR assuming perfect CSI. To obtain a formula for the average capacity with imperfect CSI, we have to consider an information theoretic approach. Regarding (\ref{EQ_MRC_SD}) and (\ref{EQ_MRC_ID}), the total signal in the MRC output at the destination node after two time slots is
\begin{equation}\label{EQ_Y_REC1}
\by_{_{o}}=\bh_o x+\bn_o,
\end{equation}
where
$\by_{_{o}}=\left[D_{_{sd}}\quad D_{_{i^\star d}}\right]^T,$
$
\bh_o=\left[\begin{array}{cc}\frac{\rho_{_{sd}}{{\mathcal{E}_{_{sd}}}}|\hat{h}_{_{sd}}|^2}{\it N_0}&\cM\end{array}\right]^T,
$
$x$ is the transmitted signal and is a scalar, and
$\bn_o=\left[\begin{array}{cc}\frac{\sqrt{{\mathcal{E}_{_{sd}}}}\hat h_{_{sd}}^*}{\it N_0}(\sqrt{{\mathcal{E}_{_{sd}}}} d_{_{sd}}x+n_{_{sd}})&\cD+\cN\end{array}\right]^T,$
where $\cD$ and $\cN$ are given in (\ref{EQ_D}) and (\ref{EQ_NN}), respectively.
Normalizing the received vector $\by_o$ by the noise variance we arrive at
\begin{equation}\label{EQ_Y_REC}
\by=\bh x+\bn,
\end{equation}
where
\begin{equation}
\by=\left[\begin{array}{cc}\frac{D_{_{sd}}}{\sqrt{{\rm var}\{D_{_{sd}}\}}}&\frac{D_{_{i^\star d}}}{\sqrt{{\rm var}\{D_{_{i^\star d}}\}}}\end{array}\right]^T,
\end{equation}
where
\begin{equation}\label{EQ_h}
\bh=\left[\begin{array}{cc}\frac{\frac{\rho_{_{sd}}{{\mathcal{E}_{_{sd}}}}|\hat{h}_{_{sd}}|^2}{\it N_0}}{\sqrt{{\rm var}\{D_{_{sd}}\}}}&\frac{\cM}{\sqrt{{\rm var}\{D_{_{i^\star d}}\}}}\end{array}\right]^T,
\end{equation}
and $\bn\sim\mathcal{CN}(\b0,\bI).$
If the noise vector $\bn$ is independent of the input signal $x$, then the capacity of the system in (\ref{EQ_Y_REC}) is given by
\begin{eqnarray}
C&=&\arg~\max_{f(x)}~\mathcal{I}(\by,x)\no\\
&=&\arg~\max_{f(x)}~\mathcal{H}(\by)-\mathcal{H}(\by|x)\no\\
&=&\arg~\max_{f(x)}~\mathcal{H}(\by)-\mathcal{H}(\bn|x)\label{EQ_capacity_info2}\\\
&=&\arg~\max_{f(x)}~\mathcal{H}(\by)-\mathcal{H}(\bn)\label{EQ_capacity_info3}\\
&=&\frac{W}{2}\log_2\left(1+\bh^H\bh\right).\label{EQ_CAP_HH}
\end{eqnarray}
where $\mathcal{H}(\cdot)$ denotes the differential entropy and $f(x)$ is the PDF of the input signal $x$.
In our system the noise vector $\bn$ depends on the input signal $x$, and on the other hand, the noise vector $\bn$ is not Gaussian. Therefore, (\ref{EQ_capacity_info2}) will not lead to (\ref{EQ_capacity_info3}) and hence we are not allowed to use the capacity formula in (\ref{EQ_CAP_HH}).
Instead, using the following inequality\cite{Cover}
\[\mathcal{H}(\bn|x)\leq\mathcal{H}(\bn),\]
we can bound the capacity with the worst effect of the additive noise on the capacity of the system \cite{Behbahani2008}, which is when the noise vector $\bn$ is Gaussian.
It has been shown in \cite{Hassibi}, Theorem 1, that if the transmitted signal and additive
noise are uncorrelated, the worst case noise has zero-mean
Gaussian distribution with the same power of the additive
noise. Furthermore, it can be easily seen from (\ref{EQ_Y_REC}) that ${\rm E}\{\bn x|\hat{h}_{_{sd}},\hat{h}_{_{si}},\hat{h}_{_{id}}\}=\b0$.
Therefore, the additive noise $\bn$ can be replaced by a Gaussian noise with the power 1, leading to the worst case capacity given by (\ref{EQ_CAP_HH}). By substituting $D_{_{sd}}$ and $D_{_{id}}$ from (\ref{EQ_MRC_SD}) and (\ref{EQ_MRC_ID}) in (\ref{EQ_h}) and constituting  (\ref{EQ_CAP_HH}), we have
\[\bh^H\bh=\gamma_r.\]}
Therefore, the worst case average capacity is given by
\[\bar{C}_{{\tt worst }}=\frac{W}{2}\int_0^\infty\log_2(1+\gamma_r)f_{\gamma_r}(\gamma_r)d\gamma_r.\]
To obtain the lower bound average capacity\footnote{$\log_2(1+\gamma)f_{\gamma}(\gamma)$ is an increasing function of $\gamma$; hence, for $\gamma_{_{ub}}\geq\gamma_{_r}$, $\int_0^{\infty}\log_2(1+\gamma_{_{ub}})f_{\gamma_{_{ub}}}(\gamma_{_{ub}})d\gamma_{_{ub}}\leq\int_0^{\infty}\log_2(1+\gamma_{_r})f_{\gamma_{_r}}(\gamma_{_r})d\gamma_{_r}.$} $\bar{C}_{_{lb}}\leq\bar{C}_{{\tt worst}}\leq\bar{C}$, we substitute $\gamma_{_r}$ with $\gamma_{_{ub}}$ and $f_{\gamma_{_r}}$ with (\ref{EQ_MAM}). Therefore, we get
\begin{eqnarray}¸\label{EQ_CAP}
\frac{\bar{C}_{_{lb}}}{W}\!\!\!&=&\!\!\!1.44 \sum_{p=1}^M\sum_{m=1}^{{\tiny \left(\!\!\!\!\begin{array}{c}M\\p\end{array}\!\!\!\!\right)}}(-1)^{(p+1)}\frac{[\Psi^{p}\bb]_m}{\bar{\mu}[\Psi^{p}\bb]_m-1}\left[\begin{array}{c}\!\!\!\bar{\mu}\exp(\frac1{\bar{\mu}})E_1(\frac1{\bar{\mu}})-\frac1{[\Psi^{p}\bb]_m}\exp([ \Psi^{p}\bb]_m)E_1([ \Psi^{p}\bb]_m)\end{array}\!\!\!\right].\no\\
\end{eqnarray}
Here, we have used the fact that \cite{Gradshteyn}
\begin{eqnarray}
\int_0^{\infty} \exp(-\alpha\gamma)\log_2(1+\gamma)d\gamma\!\!\!&=&\!\!\!-\frac{1.44}{\alpha}\exp(\alpha)Ei(-\alpha)\no\\
\!\!\!&=&\!\!\!\frac{1.44}{\alpha}\exp(\alpha)E1(\alpha) \no,
\end{eqnarray}
where  $E_1(\gamma)=-Ei(-\gamma)$, and
\begin{equation}
Ei(\gamma)\stackrel{\Delta}{=}\int_\gamma^{\infty}\frac1{x}\exp(-x)dx\no
\end{equation}
is the \emph{exponential integral function}.

\section{Asymptotic Order of Diversity}\label{SEC_Diversity}
 We shall next analyze the underlying selection cooperation system from the diversity point of view. The diversity order is given by the magnitude of the slope of the outage probability as a function of SNR (on a log-log scale) \cite{Nabar2004}. Equivalently a system achieving a diversity order $d(R)$, at transmission rate $R$, has an error probability that behaves as $\bar{P}_{_{ub}}(SNR)\varpropto SNR^{-d(R)}$ at high SNR \cite{Nabar2004}. Utilizing the high SNR analysis which is indicated in (\ref{EQ_THEORY42}), the ASER is implicitly proportional to $\bar{\gamma}_{_{sd}}, \bar{\gamma}_{_{si}}$, and $\bar{\gamma}_{_{id}}$. Assuming $\bar{\gamma}_{_{sd}}=\alpha_1\bar{\gamma}_{_{si}}=\alpha_2\bar{\gamma}_{_{id}}=P_{_{0}}$, then we have
\newcounter{mytempeqncnte10}
\normalsize \setcounter{mytempeqncnte10}{\value{equation}}
\setcounter{equation}{63}
\begin{eqnarray}\label{EQ_DIVersity}
\bar{P}_{_{ub}}(P_{_{0}})&\varpropto& \frac{1+\epsilon_{_{sd}}}{\rho_{_{sd}}^2P_{_{0}}}\prod_{i=1}^M\frac{P_{_{0}}(\alpha_2\lambda_{_{si}}+\alpha_1\lambda_{_{id}})}{P_{_{0}}^2\rho_{_{si}}^2\rho_{_{id}}^2}\no\\
&\varpropto& \frac{1+\epsilon_{_{sd}}}{\rho_{_{sd}}^2P_{_{0}}}\prod_{i=1}^M\frac{(\alpha_2\lambda_{_{si}}+\alpha_1\lambda_{_{id}})}{P_{_{0}}\rho_{_{si}}^2\rho_{_{id}}^2}\no\\
&\varpropto&\frac{1+\epsilon_{_{sd}}}{\rho_{_{sd}}^2P_{_{0}}^{M+1}}\prod_{i=1}^M\frac{(\alpha_2\lambda_{_{si}}+\alpha_1\lambda_{_{id}})}{\rho_{_{si}}^2\rho_{_{id}}^2}.
\end{eqnarray}
This implies that the diversity order in a selection cooperation scenario is highly affected by the channel estimation error. In the case that the channel estimation error is zero or it reduces by increasing the received SNR, $\epsilon_{_{sd}}, \lambda_{_{si}}, \lambda_{_{id}}, \rho_{_{sd}}, \rho_{_{si}}$, and $\rho_{_{id}}$, for $i=1, \ldots, M$, would all be equal to unity at high SNR regimes; hence,
\begin{equation}\label{EQ_div}
\bar{P}_{_{ub}}(P_{_{0}})\varpropto (\frac1{P_{_{0}}})^{M+1}.
\end{equation}
Thus, the maximum  diversity order is achieved in selection cooperation schemes with no channel estimation error or in the cases that the channel estimation error reduces with increasing SNR.

On the other hand, if the channel estimation error is independent of the received SNR, noting that $\epsilon_{_{sd}}=\frac{{\mathcal{E}_{_{sd}}}}{N_0}\sigma^2_{D,sd}$, $\epsilon_{_{si}}=\frac{{\mathcal{E}_{_{si}}}}{N_0}\sigma^2_{D,si}$, $\epsilon_{_{id}}=\frac{{\mathcal{E}_{_{id}}}}{N_0}\sigma^2_{D,id}$, $\lambda_{_{si}}=1+\rho_{_{si}}^2\epsilon_{_{id}}$, and $\lambda_{_{id}}=1+\rho_{_{id}}^2\epsilon_{_{si}}$, we can replace $\epsilon_{_{sd}}$ and similarly $\lambda_{_{si}}$, and $\lambda_{_{id}}$ by $P_{_{0}}$, at high SNR; hence, from (\ref{EQ_DIVersity}) the diversity order of the system with selection cooperation tends to zero as $P_{_{0}}$, approaches infinity. This means that in such a system we expect an error floor in the ASER or outage probability curves at high SNR, in which the performance of the system would not improve by increasing the SNR.
\section{Simulation Results}\label{SEC_SIMULATION}
{In this section, we consider two relay selection scenarios with $M=2$, and $M=3$, respectively. The transmitted symbols are drawn from the PSK constellation  where we assume $k=2$ and the received energies are assumed to be $({\mathcal{E}_{_{sd}}}, \mathcal{E}_{_{si}}, \mathcal{E}_{_{id}})=(1, 0.5, 0.5)\times P_{_{0}}$ for all relays. It is assumed that the power budget of each relay can not exceed $P_0/2$ and once a relay is selected, it operates with full power and the other relays remain silent.} All the channels are i.i.d zero mean and unit variance complex Gaussian random variables. For convenience, we  assume the special case, where the channels estimation errors of all links have the same variance, i.e., $\sigma^2_{D,sd}=\sigma^2_{D,si}=\sigma^2_{D,id}=\sigma^2_D$ and all the correlation coefficients are equal, meaning $\rho_{_{sd}}=\rho_{_{si}}=\rho_{_{id}}=\rho$. The variance of noise components is set to $N_0=1$ and $R=1$ bps/Hz. Here, unless stated, we assume that $\sigma^2_D$ is independent of the received SNR.

\begin{figure}[t]
\begin{center}  
\psfrag{P_0 }{\!\!$P_{_{0}}$}
\psfig{figure=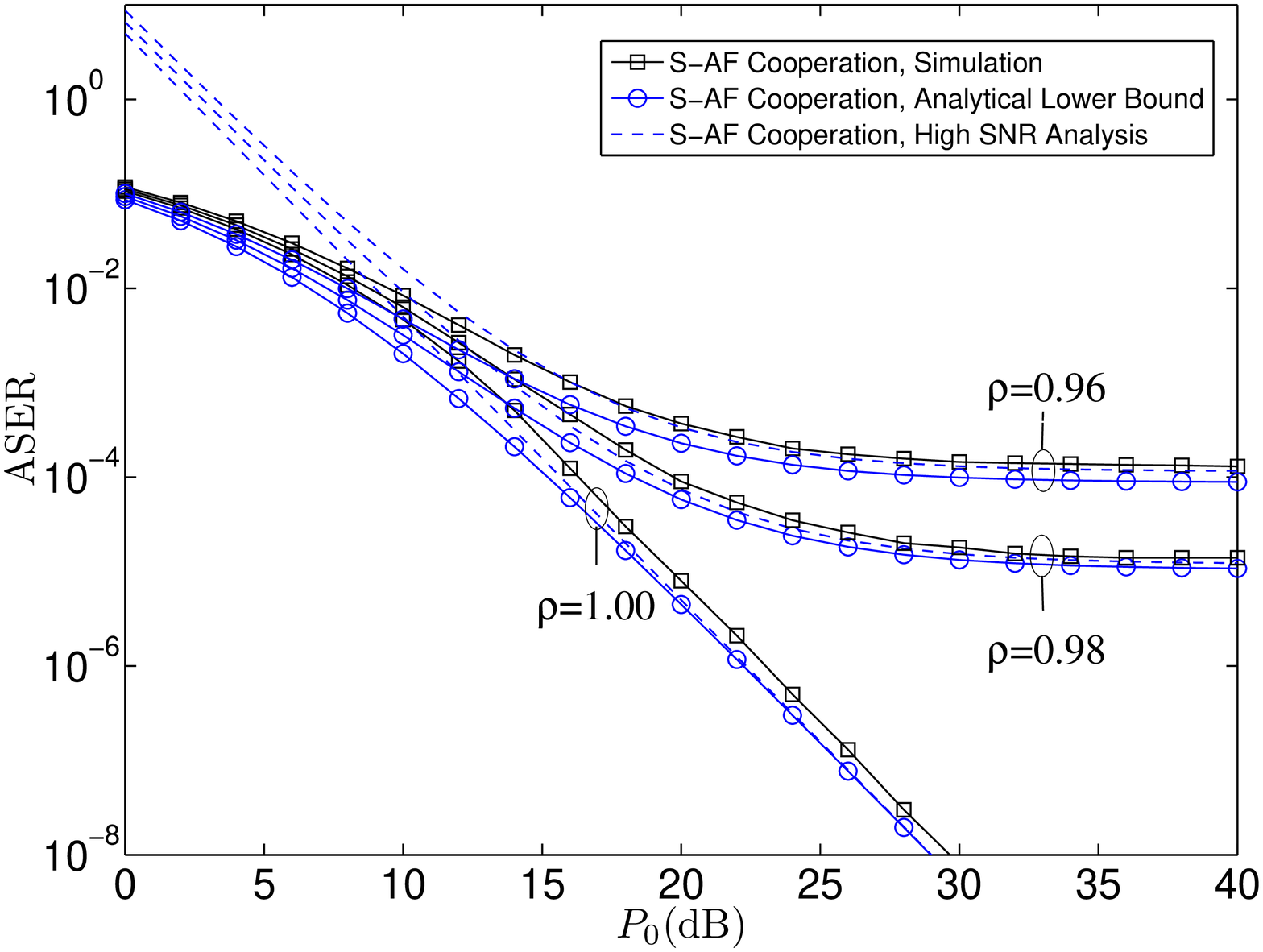,width=8in}
\caption{ASER vs. SNR. Relay selection scenario, M=2.}\label{Fig_M1}
\end{center}
\end{figure}

\begin{figure}[t]
\begin{center}  
\psfrag{P_0 }{\!\!$P_{_{0}}$}
\psfig{figure=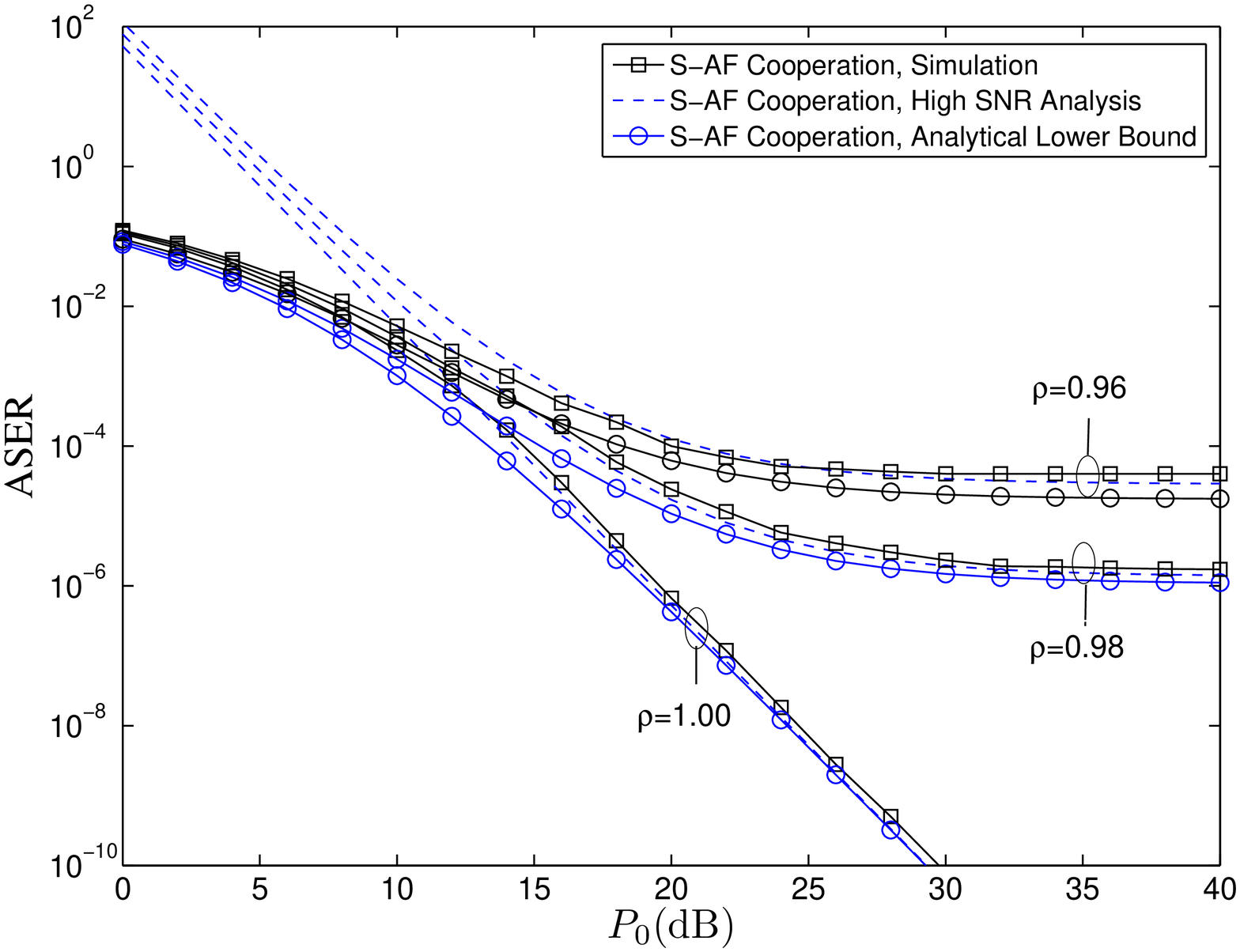,width=8in}
\caption{ASER vs. SNR. Relay selection scenario, M=3.}\label{Fig_M2}
\end{center}
\end{figure}

\emph{ASER analysis}: In Figs.~\ref{Fig_M1} and \ref{Fig_M2}, we confirm the validity of the derived expressions for the cases $M=2$, and $M=3$, respectively.
 As the figures show, the derived lower bound for the ASER is fairly tight. The high SNR analysis also converges to the analytically derived lower bound ASER
 curves at high SNR. As it is obvious in these figures, if the channel is estimated with even a very small error, an error floor is observed at high SNR.
Increasing $\rho$ yields a better performance in the ASER of the system.
Also we can deduce from these figures that selection among a larger number of relays improves the ASER performance of the system.

\begin{figure}
\begin{center}  
\psfrag{P_0 }{\!\!$P_{_{0}}$}
\psfig{figure=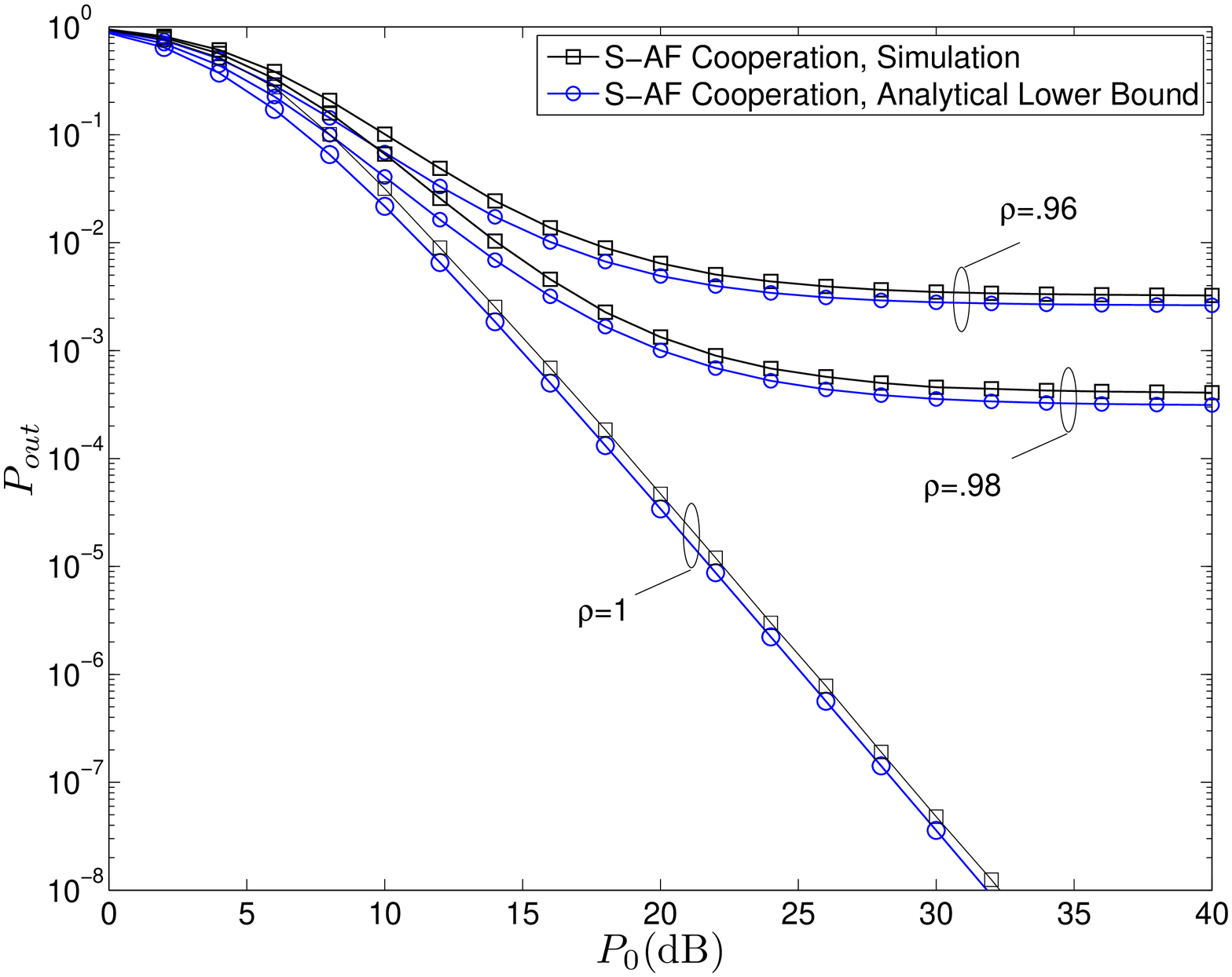,width=8in}
\caption{$P_{out}$ vs. SNR. Relay selection scenario, M=2.}\label{Fig_M3}
\end{center}
\end{figure}

\begin{figure}
\begin{center}  
\psfrag{P_0 }{\!\!$P_{_{0}}$}
\psfig{figure=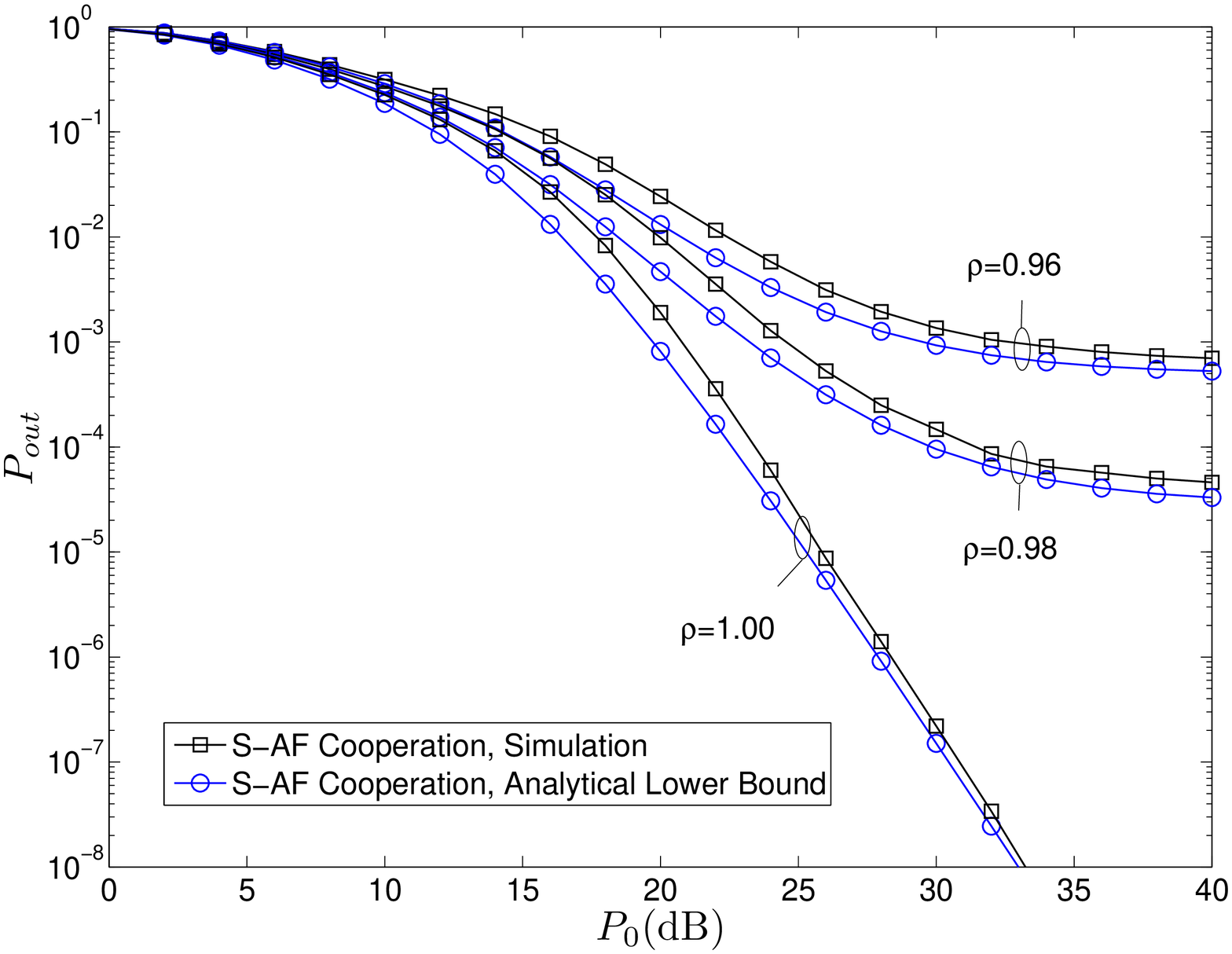,width=8in}
\caption{$P_{out}$ vs. SNR. Relay selection scenario, M=3.}\label{Fig_M4}
\end{center}
\end{figure}

\emph{Outage probability analysis}:
Figs.~\ref{Fig_M3} and \ref{Fig_M4} considers the credibility of our analytic results for $M=2$, and $M=3$. In these figures the derived formula for the outage probability is compared with the simulation results. A fair match is observed and the same error floor is seen for the case that the system faces a non-decreasing channel estimation error with the received SNR. It is observed in Figs.~\ref{Fig_M3} and \ref{Fig_M4} that the selection scheme with larger number of relays outperforms the scenario with less number of relays in terms of outage probability. Also the effect of increasing of $\rho$ and thus, decreasing of $\sigma^2_D$ in the improvement of the system outage probability is shown in these figures.
We further study the effect of relay location in selection cooperation. In this scenario the distance between $S$ and $D$ is normalized to one. The received energies are modeled as $\mathcal{E}_{_{si}}=\frac{0.5P_0}{\tilde{d}^8}$ and $\mathcal{E}_{_{id}}=\frac{0.5P_0}{(1-\tilde{d})^8}$ where $\tilde{d}$ is the normalized distance between $S$ and $R_i$.

\begin{figure}
\begin{center}  
\psfrag{P_0 }{\!\!$P_{_{0}}$}
\psfig{figure=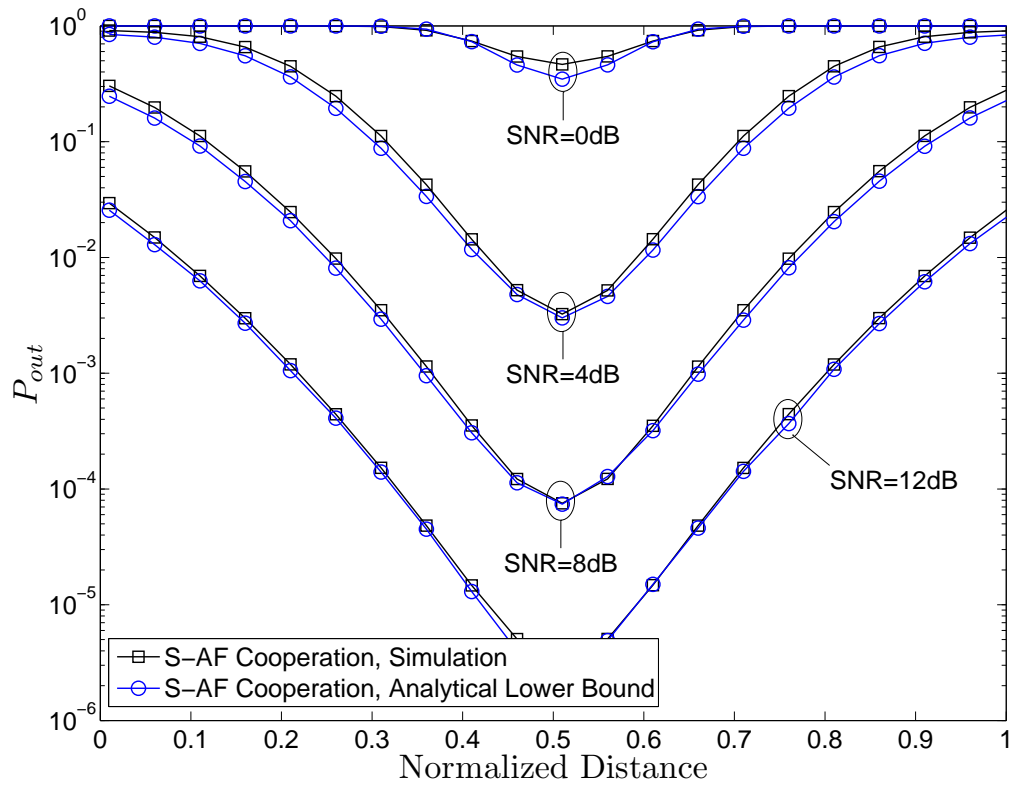,width=8in}
\caption{Outage probability for $M=2$ relays vs. distance for different SNR values.}\label{Fig_M5}
\end{center}
\end{figure}

Fig.~\ref{Fig_M5} shows the outage probability of a relay selection scenario with $M=2$, versus $\tilde{d}$. { The two relays are located between the source and destination nodes with the same distance of $\tilde{d}$ from the source node.  As it is noticed the lower bound analysis curves fairly match the curves obtained from simulation. The lowest outage probability is achieved when the selected relay is in the middle of the line of sight of source and destination.}

\emph{Asymptotic Diversity Order}:
Figs.~\ref{Fig_M1}-\ref{Fig_M4} also considers the credibility of our diversity analysis for $M=2$ and $M=3$.
The slope of the curves associated with $\rho=1$ in these figures imply the diversity orders achieved by the selection cooperation scenario.
Diversity order $M+1$ is seen for the perfect channel estimation scenario. Occurrence of error floor in the curves associated with the imperfect channel estimation accounts for the diversity order of zero at high SNR's which is analytically obtained.

\begin{figure}
\begin{center}  
\psfrag{P_0 }{\!\!$P_{_{0}}$}
\psfig{figure=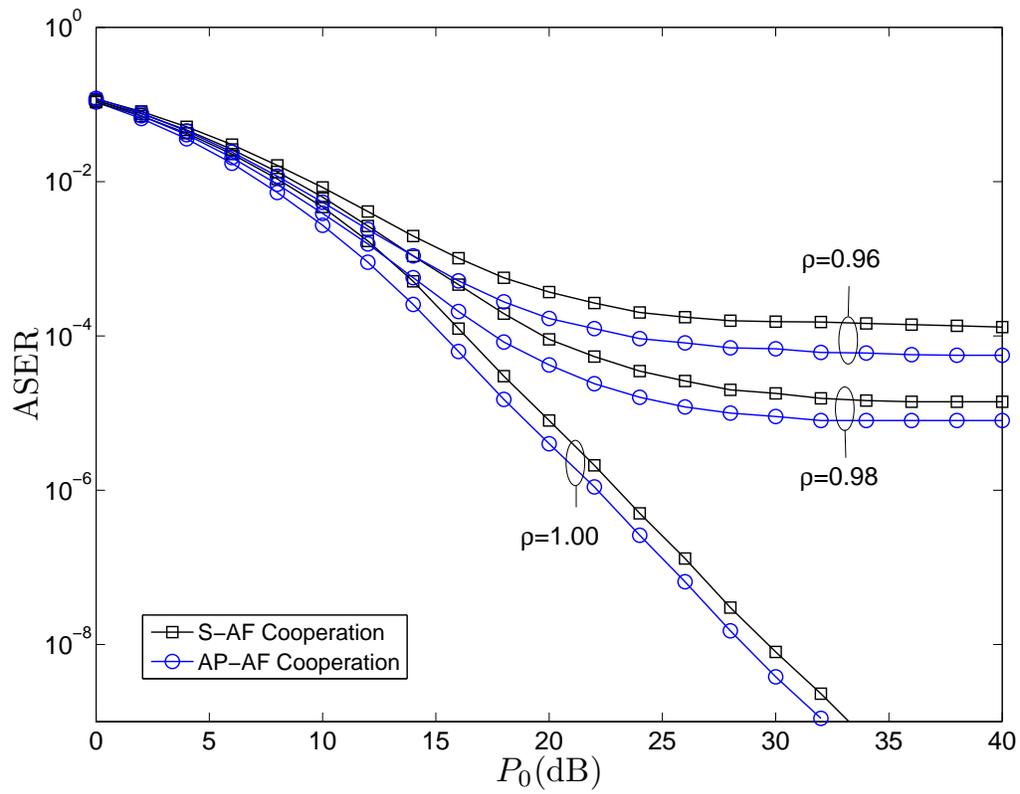,width=8in}
\caption{ASER vs SNR. Performance of S-AF and AP-AF Cooperation for M=2.}\label{Fig_M7}
\end{center}
\end{figure}

  Fig.~\ref{Fig_M7} shows the ASER performance curves for S-AF and all participate amplify-and-forward cooperation (AP-AF), where all the relays participate in cooperation. It is noticed that both scenarios exhibit the same slope, i.e., same diversity order. This means that S-AF as well as AP-AF in the cooperative system achieve full diversity order, i.e,  $M+1$, where $M$ is the number of relays. Although selection degrades the ASER performance of the system, it improves the throughput of the system significantly. Also S-AF cooperation requires less power consumption than the AP-AF cooperation scheme.

\begin{figure}
\begin{center}  
\psfrag{P_0 }{\!\!$P_{_{0}}$}
\psfig{figure=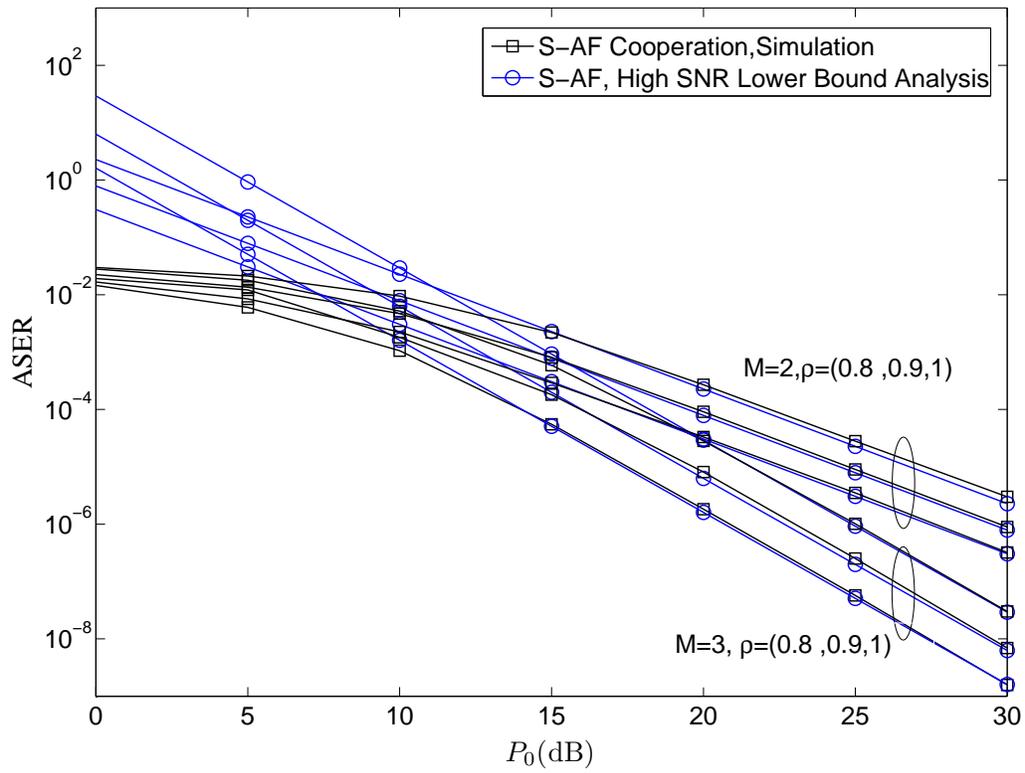,width=8in}
\caption{ASER Versus SNR. S-AF for M=2 and 3, when the estimation error variance decreases with increasing the SNR.}\label{Fig_M8}
\end{center}
\end{figure}

The same results are achievable for the case that the estimation error variance is decreased by increasing the SNR. Fig.~\ref{Fig_M8} shows the simulation results for $M=2$ and $M=3$ simultaneously. In this figure the received energies are assumed to be $({\mathcal{E}_{_{sd}}}, \mathcal{E}_{_{si}}, \mathcal{E}_{_{id}})=(0.5, 0.5, 0.5)\times P_{_{0}}$, for all relays. Diversity orders of 3 and 4 are achieved for $M=2$ and $M=3$, respectively.

\begin{figure}
\begin{center}  
\psfrag{P_0 }{\!\!$P_{_{0}}$}
\psfig{figure=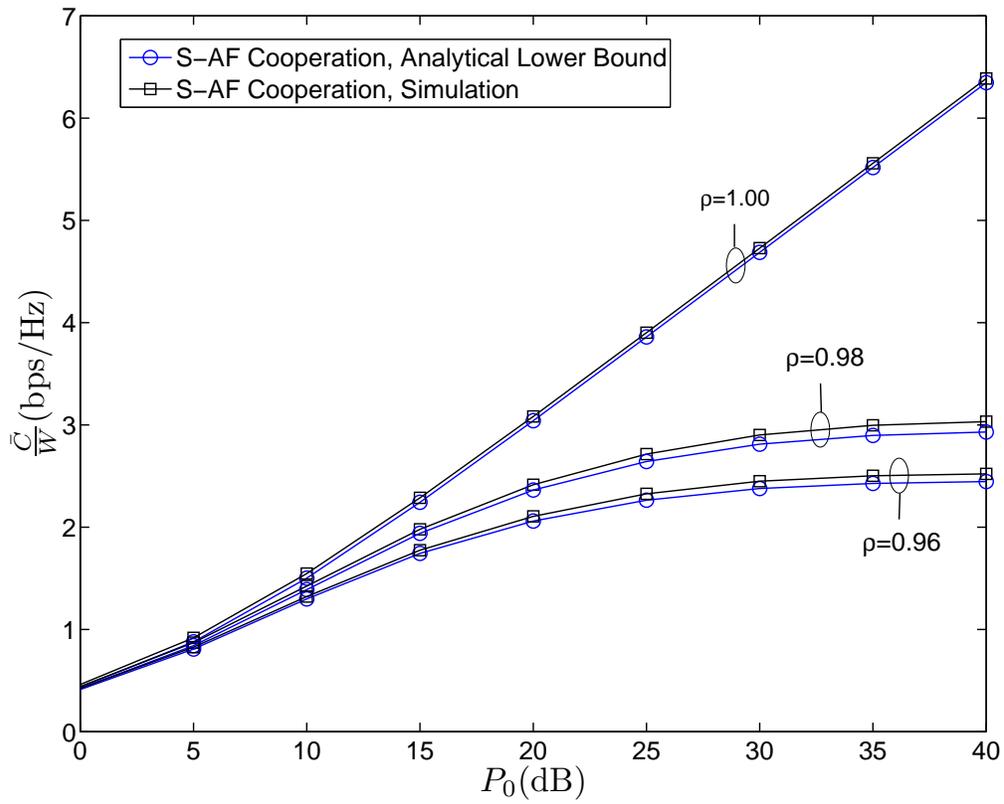,width=8in}
\caption{Average capacity per bandwidth vs. SNR. Relay selection Scenario, M=2.}\label{Fig_M9}
\end{center}
\end{figure}

\begin{figure}
\begin{center}  
\psfrag{P_0 }{\!\!$P_{_{0}}$}
\psfig{figure=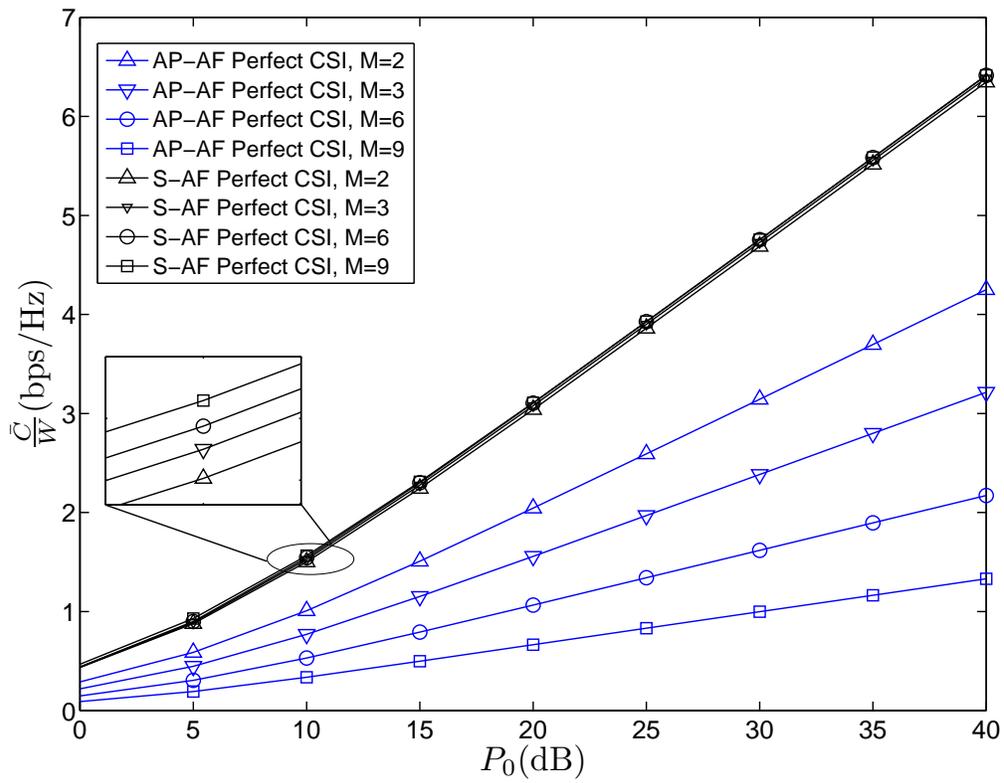,width=8in}
\caption{Average capacity per bandwidth vs. SNR. Comparison between S-AF and AP-AF cooperation schemes.}\label{Fig_M10}
\end{center}
\end{figure}

\emph{Average Capacity}:
Fig.~\ref{Fig_M9} shows the capacity of the S-AF scenario when we select the best relay out of $M=2$, relays. A capacity ceiling is observed when the channel is estimated with error. After the effect of the receiver noise is completely diminished, which occurs at a certain SNR, the capacity improvement is insignificant from that point on. This is in fact because increasing the SNR does not reduce the channel estimation error in our scenario. As it can be seen, the capacity of the system is very sensitive to the channel estimation performance. A fair upper bound is observed in our simulations. Furthermore, Fig.~\ref{Fig_M10} shows a comparison between S-AF and AP-AF cooperation for $M=2,3,6$, and $9$. Obviously, S-AF cooperation exhibits a huge average capacity improvement specially for $m>2$. Selection out of larger number of relays would slightly improve the throughput of the system.

\section{Conclusion}\label{SEC_CONCLUSION}
In this paper we study the effect of channel estimation error in cooperative systems with relay selection. We analyze the ASER and outage probability performance the system in the presence of estimation error. We further derive closed form formulations for the average capacity of the selection cooperation with imperfect channel estimates. Our analysis shows that the selection cooperation scheme still preserves the full diversity order. A simulation study is presented to corroborate the analytical
results and to demonstrate the performance of relay selection with imperfect channel estimation.

\section*{Appendix.~I\\Deriving the Probability density function of $\gamma_{_{i^{{\tiny\star}}\!\!,ub}}$}
Consider a case with three relays, i.e., $M=3$. From {\emph{Theorem 3}} we have
\begin{equation}
F_{\gamma_{_{i^{{\tiny\star}}\!\!,ub}}}(\gamma)=\prod_{i=1}^3(1-\exp(-\frac{\gamma}{\beta_i})).\no
\end{equation}
Using (\ref{EQ_BETTA_i}) to (\ref{EQ_theta_ub}), we have
{\small
\begin{eqnarray}
f_{\gamma_{_{i^{{\tiny\star}}\!\!,ub}}}(\gamma)&=&\sum_{l=1}^3\frac1{\beta_l}\exp(-\frac{\gamma}{\beta_l})\prod_{{\tiny\begin{array}{c}i=1\\i\neq l\end{array}}}^3(1-\exp(-\frac{\gamma}{\beta_i}))\no\\
&=&\frac1{\beta_1}\left[\begin{array}{c}\exp(-\frac{\gamma}{\beta_1})-\exp(-\gamma(\frac1{\beta_1}+\frac1{\beta_2}))-\exp(-\gamma(\frac1{\beta_1}+\frac1{\beta_3}))-\exp(-\gamma(\frac1{\beta_1}+\frac1{\beta_2}+\frac1{\beta_3}))\end{array}\right]\no\\
&+&\frac1{\beta_2}\left[\begin{array}{c}\exp(-\frac{\gamma}{\beta_2})-\exp(-\gamma(\frac1{\beta_1}+\frac1{\beta_2}))-\exp(-\gamma(\frac1{\beta_2}+\frac1{\beta_3}))-\exp(-\gamma(\frac1{\beta_1}+\frac1{\beta_2}+\frac1{\beta_3}))\end{array}\right]\no\\
&+&\frac1{\beta_3}\left[\begin{array}{c}\exp(-\frac{\gamma}{\beta_3})-\exp(-\gamma(\frac1{\beta_1}+\frac1{\beta_3}))-\exp(-\gamma(\frac1{\beta_2}+\frac1{\beta_3}))-\exp(-\gamma(\frac1{\beta_1}+\frac1{\beta_2}+\frac1{\beta_3}))\end{array}\right]\no\\
&=&\frac1{\beta_1}\exp(-\frac{\gamma}{\beta_1})+\frac1{\beta_2}\exp(-\frac{\gamma}{\beta_2})+\frac1{\beta_3}\exp(-\frac{\gamma}{\beta_3})\no\\
&-&(\frac1{\beta_1}+\frac1{\beta_2})\exp(-\gamma(\frac1{\beta_1}+\frac1{\beta_2}))-(\frac1{\beta_1}+\frac1{\beta_3})\exp(-\gamma(\frac1{\beta_1}+\frac1{\beta_3}))-(\frac1{\beta_2}+\frac1{\beta_3})\exp(-\gamma(\frac1{\beta_2}+\frac1{\beta_3}))\no\\
&+&(\frac1{\beta_1}+\frac1{\beta_2}+\frac1{\beta_3})\exp(-\gamma(\frac1{\beta_1}+\frac1{\beta_2}+\frac1{\beta_3})).\label{EQ_EX}
\end{eqnarray}}

Let $\Psi^{p}$ be a binary permutation matrix with dimension of ${{\tiny\left(\!\!\!\begin{array}{c}M\\p\end{array}\!\!\!\right)\times M}}$. Each row shows one possible $p$ combination of $M$ binary bits.
In this example
\begin{eqnarray}
\Psi^1=\left[\begin{array}{ccc}1&0&0\\0&1&0\\0&0&1\end{array}\right],\Psi^2=\left[\begin{array}{ccc}1&1&0\\0&1&1\\1&0&1\end{array}\right],\Psi^3=\left[\begin{array}{ccc}1&1&1\end{array}\right].\no
\end{eqnarray}
We also define $\bb=\left[\begin{array}{ccc}\frac1{\beta_1}&\frac1{\beta_2}&\frac1{\beta_3}\end{array}\right]^T$ and let $[\Psi^p\bb]_m$ be the $m^{th}$ row of the vector $\Psi^p\bb$. Using the definitions we just pointed out we can write (\ref{EQ_EX}) as
\begin{equation}
 f_{\gamma_{_{i^{{\tiny\star}}\!\!,ub}}}(\gamma)=\sum_{p=1}^3\sum_{m=1}^{{\tiny \left(\!\!\!\!\begin{array}{c}3\\p\end{array}\!\!\!\!\right)}}(-1)^{(p+1)}[\Psi^{p}\bb]_m\exp\left(\begin{array}{c}\!\!\!-\gamma[\Psi^{p}\bb]_m\!\!\!\end{array}\right).
\end{equation}
With the same procedure for $M$ relays, using (\ref{EQ_BETTA_i}) and substituting  $F_{\gamma_{_{i,ub}}} = [1-\exp(-\frac{\gamma}{\beta_i})]$ into (\ref{EQ_theta_ub}), we obtain
\begin{equation}\label{EQ_psi}
 f_{\gamma_{_{i^{{\tiny\star}}\!\!,ub}}}(\gamma)=\sum_{p=1}^M\sum_{m=1}^{{\tiny \left(\!\!\!\!\begin{array}{c}M\\p\end{array}\!\!\!\!\right)}}(-1)^{(p+1)}[\Psi^{p}\bb]_m\exp\left(\begin{array}{c}\!\!\!-\gamma[\Psi^{p}\bb]_m\!\!\!\end{array}\right),
\end{equation}
where $\bb$ is defined in (\ref{EQ_b}).

  Furthermore,  it can be easily shown that the PDF of $\mu=\frac{\rho_{_{sd}}^2\hat{\gamma}_{_{sd}}}{1+\epsilon_{_{sd}}}$ is given by $f_{\mu}(\gamma)=\frac1{\bar{\mu}}\exp(-\frac{\gamma}{\bar{\mu}})$, where $\bar{\mu}=\frac{\rho_{_{sd}}^2\bar{\gamma}_{_{sd}}}{1+\epsilon_{_{sd}}}$, and the PDF of $\gamma_{_{ub}}$, can be found as follows

\begin{eqnarray}
f_{\gamma_{_{ub}}}(\gamma)\!\!\!&=&\!\!\!\frac{\partial}{\partial\gamma}P(\gamma_{_{ub}}\leq\gamma)\no\\
\!\!\!&=&\!\!\!\frac{\partial}{\partial\gamma}P(\mu+\gamma_{_{i^{{\tiny\star}}\!\!,ub}}\leq\gamma)\no\\
\!\!\!&=&\!\!\!\frac{\partial}{\partial\gamma}\int_0^\gamma P(\mu\leq\gamma-\gamma_{_{i^{{\tiny\star}}\!\!,ub}})f_{\gamma_{_{i^{{\tiny\star}}\!\!,ub}}}d\gamma_{_{i^{{\tiny\star}}\!\!,ub}}\no\\
\!\!\!&=&\!\!\!\frac{\partial}{\partial\gamma}\int_0^\gamma F_{\mu}(\gamma-\gamma_{_{i^{{\tiny\star}}\!\!,ub}})f_{\gamma_{_{i^{{\tiny\star}}\!\!,ub}}}d\gamma_{_{i^{{\tiny\star}}\!\!,ub}}.\label{EQ_last_EQ}
\end{eqnarray}
Using $F_\mu(\gamma)=1-\exp(-\frac{\gamma}{\bar{\mu}})$ and inserting (\ref{EQ_psi}) in (\ref{EQ_last_EQ}), we obtain
\begin{eqnarray}
f_{\gamma_{_{ub}}}(\gamma)\!\!\!&=&\!\!\!\sum_{p=1}^M\sum_{m=1}^{{\tiny \left(\!\!\!\!\begin{array}{c}M\\p\end{array}\!\!\!\!\right)}}(-1)^{(p+1)}\frac{\partial}{\partial\gamma}\int_0^\gamma[1-\exp(-\frac{\gamma-\gamma_{_{i^{{\tiny\star}}\!\!,ub}}}{\bar{\mu}})][\Psi^{p}\bb]_m\exp\left(\begin{array}{c}\!\!\!-\gamma_{_{i^{{\tiny\star}}\!\!,ub}}[\Psi^{p}\bb]_m\!\!\!\end{array}\right)d{\gamma_{_{i^{{\tiny\star}}\!\!,ub}}}\no\\
\!\!\!&=&\!\!\!\sum_{p=1}^M\sum_{m=1}^{{\tiny \left(\!\!\!\!\begin{array}{c}M\\p\end{array}\!\!\!\!\right)}}(-1)^{(p+1)}\frac{[\Psi^{p}\bb]_m}{\bar{\mu}[\Psi^{p}\bb]_m-1}\left[\begin{array}{c}\!\!\!\exp(\frac{-\gamma}{\bar{\mu}})-\exp(-\gamma[ \Psi^{p}\bb]_m)\end{array}\!\!\!\right].
\end{eqnarray}

\section*{Appendix.~II\\ Proof of {\it Theorem 4}}
From the basic principles of the moment generation function (MGF) of $\gamma_{_{ub}}=\mu+\gamma_{_{i^{{\tiny\star}}\!\!,ub}}$, we have
\begin{equation}
M_{\gamma_{_{ub}}}(s)=M_{\mu}(s)M_{\gamma_{_{i^{{\tiny\star}}\!\!,ub}}}(s),
\end{equation}
thus using the initial value theorem, the value of $\frac{\partial^nf_{\gamma_{_{ub}}}}{\partial\gamma_{_{ub}}^n}(0)$ is given by
\begin{eqnarray}
\frac{\partial^{n}f_{\gamma_{_{ub}}}}{\partial\gamma_{_{ub}}^{n}}(0)&=&\lim_{s\rightarrow \infty}s^{n+1}M_{\gamma_{_{ub}}}(s)\no\\
&=&\lim_{s\rightarrow \infty}sM_{\mu}(s)s^{n}M_{\gamma_{_{i^{{\tiny\star}}\!\!,ub}}}(s)\no\\
&=&f_{\mu}(0)\frac{\partial^{n-1}f_{\gamma_{_{i^{{\tiny\star}}\!\!,ub}}}}{\partial\gamma_{_{i^{{\tiny\star}}\!\!,ub}}^{n-1}}(0),\label{EQ_MGFTHETA}
\end{eqnarray}
where
\begin{equation}
\frac{\partial^{n-1}f_{\gamma_{_{i^{{\tiny\star}}\!\!,ub}}}}{\partial\gamma_{_{i^{{\tiny\star}}\!\!,ub}}^{n-1}}(0)=\frac{\partial^{n}F_{\gamma_{_{i^{{\tiny\star}}\!\!,ub}}}}{\partial\gamma_{_{i^{{\tiny\star}}\!\!,ub}}^{n}}(0).\label{EQ_fF}
\end{equation}
In light of \emph{Theorem 3}, we can simply deduce that for $n<M$, $\frac{\partial^{n}F_{\gamma_{_{i^{{\tiny\star}}\!\!,ub}}}}{\partial\gamma_{_{i^{{\tiny\star}}\!\!,ub}}^{n}}(0)$ is the summation of some terms in the product form, where each product form contains at least one cumulative density function (CDF) of $\gamma_{i}^{ub}$, at zero. Since $F_{\gamma_{_{i,ub}}}(0)=0$, we have $\frac{\partial^{n}F_{\gamma_{_{i^{{\tiny\star}}\!\!,ub}}}}{\partial\gamma_{_{i^{{\tiny\star}}\!\!,ub}}^{n}}(0)=0$; hence, from (\ref{EQ_fF}), $\frac{\partial^{n}f_{\gamma_{_{ub}}}}{\partial\gamma_{_{ub}}^{n}}(0)=0$. For $n=M$, we have
\begin{eqnarray}
\frac{\partial^{M-1}f_{\gamma_{_{i^{{\tiny\star}}\!\!,ub}}}}{\partial\gamma_{_{i^{{\tiny\star}}\!\!,ub}}^{M-1}}(0)&=&\frac{\partial^{M}F_{\gamma_{_{i^{{\tiny\star}}\!\!,ub}}}}{\partial\gamma_{_{i^{{\tiny\star}}\!\!,ub}}^{M}}(0)\no\\
&=&M!\prod_{i=1}^{M}f_{\gamma_{_i}}^{ub}(0)+e(\gamma_{_{i^{{\tiny\star}}\!\!,ub}}),\label{EQ_FINAL_MGF}
\end{eqnarray}
where in this equation $e(\gamma_{_{i^{{\tiny\star}}\!\!,ub}})$ denotes all the other terms. Note that each term in $e(\gamma_{_{i^{{\tiny\star}}\!\!,ub}})$ has a product form and contains at least one CDF of $\gamma_{i}^{ub}$ at zero and $e(\gamma_{_{i^{{\tiny\star}}\!\!,ub}})=0$.
Substituting (\ref{EQ_FINAL_MGF}) into (\ref{EQ_MGFTHETA}) yields
\begin{eqnarray}
\frac{\partial^{M}f_{\gamma_{_{ub}}}}{\partial\gamma_{_{ub}}^{M}}(0)&=&M!f_{\mu}(0)\prod_{i=1}^{M}f_{\gamma_{_i}}^{ub}(0)\no\\
&=&M!\frac1{\mu}\prod_{i=1}^{M}\frac1{\beta_i}.\label{EQ_Final_Derivation_ASSYMPTOTIC}
\end{eqnarray}
{By (\ref{EQ_uub}), (\ref{EQ_llb}) and \emph{Theorem 1} and following similar steps to above, we obtain
\begin{eqnarray}
\frac{\partial^{M}f_{\gamma_{_{lb}}}}{\partial\gamma_{_{lb}}^{M}}(0)=2M!\frac1{\mu}\prod_{i=1}^{M}\frac1{\beta_i}.\label{EQ_Final_Derivation_ASSYMPTOTIC2}
\end{eqnarray}}

Finally by substituting (\ref{EQ_Final_Derivation_ASSYMPTOTIC}) and (\ref{EQ_Final_Derivation_ASSYMPTOTIC2}) into (\ref{EQ_ass_P_E}) and (\ref{EQ_ass_P_El}), we obtain  the result in (\ref{EQ_THEORY4}), which completes the proof.

\hspace{95mm}~~~~~~~~~~~~~~~~~~~~~~~~~~~~~~~~~~~~~~~~~~~~~~~~~~~~~~~~~~~~~~~$\blacksquare$

\bibliographystyle{IEEEtran}
\bibliography{JOURNAL_SELECTION_DRAFT}

\newpage
\begin{figure}[center]
\begin{center}  
\psfig{figure=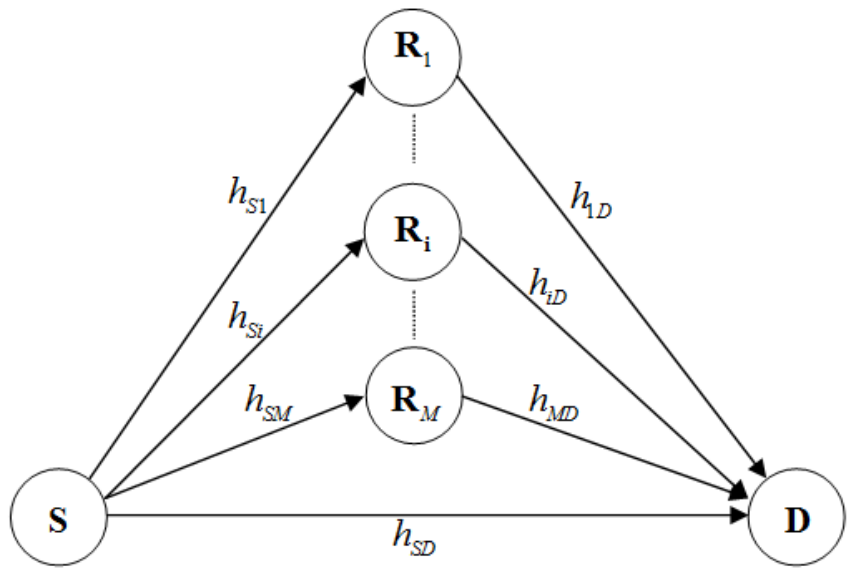,width=6in}
\caption{Schematic of cooperative communications system with M relays. Only the best relay, $R_{i^{{\tiny\star}}}$, is selected  among $R_i$s, $i=1, \ldots M$.}\label{Fig_LMS1}
\end{center}
\end{figure}

\newpage

\newpage
\newpage
\newpage
\newpage

\newpage
\newpage
\newpage
\newpage

\end{document}